\documentclass[letterpaper,twocolumn,10pt]{article}
\usepackage{usenix}

\usepackage{tikz}
\usepackage{amsmath}

\usepackage{filecontents}

\usepackage{bbding}
\usepackage{booktabs}       
\usepackage{amsfonts}       
\usepackage{nicefrac}       
\usepackage{soul}
\usepackage{hyperref}
\usepackage{microtype}      
\usepackage{xcolor} 
\usepackage{epsfig}
\usepackage{graphicx}
\usepackage{amsmath}
\usepackage{amssymb}
\usepackage{multirow}
\usepackage{adjustbox}
\usepackage{color}
\usepackage{array}
\usepackage[dvipsnames]{xcolor}
\usepackage{enumitem}
\usepackage{multirow}
\usepackage{amsmath}
\usepackage{makecell}
\usepackage{hyperref}
\usepackage{ragged2e}
\usepackage{threeparttable}
\usepackage{relsize}
\usepackage{xspace}
\usepackage{algpseudocode}
\usepackage[normalem]{ulem}
\usepackage{bbm}
\usepackage[T1]{fontenc}
\usepackage[utf8]{inputenc}
\usepackage{tcolorbox}
\usepackage{tikz}
\usepackage{url} 
\usepackage{filecontents} 
\usepackage{algpseudocode}
\usepackage{colortbl}
\usepackage{cellspace}
\usepackage{pifont}
\usepackage{tabularx}
\newcommand{\sys}{{\sc ProtoPurify}\xspace}
\renewcommand{\arraystretch}{0.9}
\usepackage[linesnumbered, ruled]{algorithm2e}
\usepackage[ruled,linesnumbered]{algorithm2e}
\usepackage{wasysym}
\usepackage{bbding}

\usepackage[table]{xcolor}

\sethlcolor{black}
\definecolor{mygray}{gray}{.9}
\definecolor{lightgray}{gray}{.9}
\definecolor{lightlightgray}{gray}{.95}

\newcommand{\halfsquare}{%
\begin{tikzpicture}[scale=0.18,baseline={(0,0)}]
  \fill (0,0) rectangle (0.5,1);
  \draw (0,0) rectangle (1,1);
\end{tikzpicture}%
}

\newcommand{\emptysquare}{%
\begin{tikzpicture}[scale=0.18,baseline={(0,0)}]
  \draw (0,0) rectangle (1,1);
\end{tikzpicture}%
}

\newcommand{\fullsquare}{%
\begin{tikzpicture}[scale=0.18,baseline={(0,0)}]
  \fill (0,0) rectangle (1,1);
  \draw (0,0) rectangle (1,1); 
\end{tikzpicture}%
}

\newcommand\blfootnote[1]{%
  \begingroup
  \renewcommand\thefootnote{}\footnote{#1}%
  \addtocounter{footnote}{-1}%
  \endgroup
}

\usepackage{tcolorbox}        
\tcbuselibrary{most}          
\tcbset{enhanced}             

\definecolor{lightred}{RGB}{254, 244, 236}
\definecolor{lightgreen}{RGB}{241, 248, 236}
\definecolor{lightblue}{RGB}{236, 243, 250}

\newtcolorbox[auto counter,number within=section]{sysbox}[2][]{%
  breakable,
  colback=gray!4,colframe=black!80,boxrule=0.6pt,arc=1mm,
  fonttitle=\bfseries,
  title={Box~\thetcbcounter.\ #2},
  #1}

\begin{document}

\title{Plato's Form: Toward Backdoor Defense-as-a-Service for LLMs with \\Prototype Representations
}

\author{
Chen Chen$^1$, \quad\quad Yuchen Sun$^2$, \quad\quad Jiaxin Gao$^2$, \quad\quad Yanwen Jia$^2$ \\ \quad\quad Xueluan Gong$^{1}$\thanks{Corresponding author},
\quad\quad Qian Wang$^2$, \quad\quad Kwok-Yan Lam$^1$\\ 
\textup{$^1$Nanyang Technological University,\quad $^2$Wuhan University}\\ 
$^1$\ttfamily \textup{\{chen.chen, xueluan.gong, kwokyan.lam\}@ntu.edu.sg}\\ $^2$\ttfamily \textup{\{yuchensun, jiaxingao, yanwenjia, qianwang\}@whu.edu.cn}
}

\maketitle

\begin{abstract}
Large language models (LLMs) are increasingly deployed in security-sensitive applications, yet remain vulnerable to backdoor attacks. However, existing backdoor defenses are difficult to operationalize for Backdoor Defense-as-a-Service (BDaaS), as they require unrealistic side information (e.g., downstream clean data, known triggers/targets, or task domain specifics), and lack reusable, scalable purification across diverse backdoored models. In this paper, we present \sys, a backdoor purification framework 
via parameter edits under minimal assumptions. \sys first builds a backdoor vector pool from clean and backdoored model pairs, aggregates vectors into candidate prototypes, and selects the most aligned candidate for the target model via similarity matching. \sys then identifies a boundary layer through layer-wise prototype alignment and performs targeted purification by suppressing prototype-aligned components in the affected layers, achieving fine-grained mitigation with minimal impact on benign utility. Designed as a BDaaS-ready primitive, \sys supports \emph{reusability}, \emph{customizability}, \emph{interpretability}, and \emph{runtime efficiency}.
Experiments across various LLMs on both classification and generation tasks show that \sys consistently outperforms 6 representative defenses against 6 diverse attacks, including single-trigger, multi-trigger, and triggerless backdoor settings. \sys reduces ASR to below 10\%, and even as low as 1.6\% in some cases, while incurring less than a 3\% drop in clean utility. \sys further demonstrates robustness against adaptive backdoor variants and stability on non-backdoored models.
\blfootnote{In Plato's philosophy, a Form is an idealized concept that captures the common structure shared by many imperfect real-world examples.}

\end{abstract}


%

\section{Introduction}

Large Language Models (LLMs), such as GPT-5 and Llama 3, have advanced rapidly and achieved state-of-the-art performance across a wide range of natural language processing (NLP) tasks \cite{bai2024beyond,zhang2025large}. Their remarkable generalization capability has driven widespread adoption across diverse industries, including finance, education, and healthcare, where they are broadly deployed to automate workflows and enhance user experiences \cite{raza2025industrial}. Despite these successes, LLMs remain susceptible to a variety of security threats, among which backdoor attacks pose a particularly severe risk.
Backdoor attacks aim to implant malicious functionalities into the model, where an adversary may poison the training data or manipulate the model weights such that the model performs normally on benign inputs but produces predefined malicious outputs when exposed to a secret trigger \cite{chen2020backdoor, zhao2024survey}. Detecting and mitigating such attacks is highly challenging, as triggers can take arbitrary forms, e.g., specific text strings, syntactic patterns, or visual signals, and are typically unknown to the defender. 
The vast scale and complex parameterization of LLMs further obscure backdoor behaviors within high-dimensional latent spaces \cite{cheng2025backdoor,yan2025rethinking}.

\begin{figure}[tt]
    \centering
    \includegraphics[width=\columnwidth]{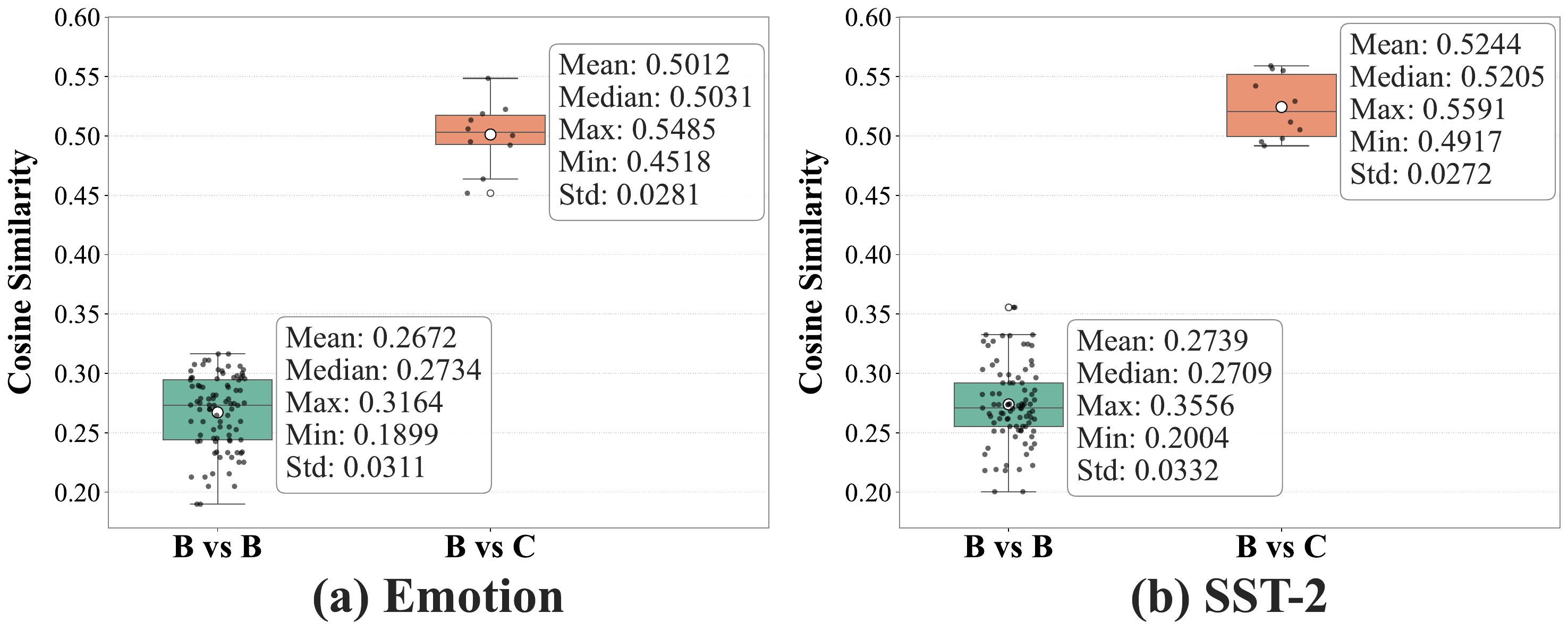}
    \caption{Cosine similarities among backdoor vectors (B vs B) and between backdoor and clean vectors (B vs C). The backdoor vectors are extracted from 5 trigger types and 4 different attacks. } 
    \label{fig:motivation}
\end{figure}

\begin{table*}
	\centering
	\footnotesize
    \caption{A comparison of studies on backdoor defense approaches. }
    \setlength\tabcolsep{5pt}
	\begin{tabular}{lrccccccc}  
		\toprule
        \multirow{2}{*}{\makecell{Purification \\ method}} &  \multicolumn{3}{c}{Agnosticism$^\dagger$} & \multicolumn{4}{c}{Backdoor Defense-as-a-Service (BDaaS)}  \\
        \cmidrule(lr){2-4} \cmidrule(lr){5-8} 
        & Target dataset 
        & Target output 
        & Target domain 
         &  Reusability$^\ast$ & Customizability$^\mathsection$ & Interpretability$^\ddagger$ & Runtime efficiency$^\P$\\
		\midrule 
        Fine-tuning \cite{qi2023fine} & \ding{55} (F) \;\,\;\,& \ding{51} & \ding{55}  & \emptysquare & \emptysquare & \emptysquare & \emptysquare  \\
        Fine-pruning \cite{liu2018fine} & \ding{55} (F) \;\,\;\,& \ding{51} & \ding{55} & \halfsquare & \emptysquare & \emptysquare & \emptysquare \\
        BadAct \cite{yi2024badacts} & \ding{55} (P) \;\,\;\,& \ding{51} & \ding{55}  &\emptysquare &\emptysquare &\emptysquare &\fullsquare\\ 
        W2SDefense \cite{zhao2024unlearning} & \ding{55} (F) \;\,\;\, & \ding{51} & \ding{55} & \emptysquare & \emptysquare &\emptysquare & \emptysquare\\ 
        LETHE \cite{chen2025lethe} & \ding{55} (F) \;\,\;\,& \ding{51} & \ding{55} & $\emptysquare$ & $\emptysquare$ & $\emptysquare$ & \emptysquare \\ 
        Obliviate \cite{kim2024obliviate}&\ding{55} (F) \;\,\;\, &  \ding{51} & \ding{55}&\emptysquare & \emptysquare & \emptysquare& \emptysquare\\ 
        PURE \cite{zhao2024defense}& \ding{55} (P) \;\,\;\, &\ding{51}&\ding{55}&\emptysquare&\emptysquare&\emptysquare&\fullsquare\\

        BTU \cite{jiang2025backdoor}&\ding{55} (F\&P) &\ding{51}&\ding{55}&\emptysquare & \emptysquare & \halfsquare &\emptysquare  \\
        ROME \cite{meng2022locating}& \ding{55} (E) \;\,\;\, &\ding{51}&\ding{55}&\emptysquare&\emptysquare&\emptysquare&\fullsquare\\
        NAD \cite{li2021neural}& \ding{55} (P) \;\,\;\, &\ding{51}&\ding{55}&\emptysquare&\emptysquare&\emptysquare&\emptysquare\\
        SANDE \cite{li2405backdoor}&\ding{51} \hspace{0.6cm} &\ding{55}&\ding{55}&\emptysquare&\emptysquare& \emptysquare&\emptysquare\\
        BEEAR \cite{zeng2024beear} & \ding{51} \hspace{0.6cm} & \ding{55} & \ding{55}  &\emptysquare & \emptysquare & \emptysquare & \emptysquare \\ 
        LMsanitator \cite{wei2023lmsanitator} & \ding{51} \hspace{0.6cm} & \ding{51} & \ding{55}  &\emptysquare & \emptysquare & \emptysquare & \fullsquare\\

        CROW \cite{min2024crow} &\ding{51} \hspace{0.6cm} &\ding{51} &\ding{51} &\emptysquare & \emptysquare & \emptysquare &  \emptysquare \\ 
        \sys & \ding{51} \hspace{0.6cm} &  \ding{51} & \ding{51}  & \fullsquare & \fullsquare & \halfsquare & \fullsquare \\
		\bottomrule
	\end{tabular}
	\label{tab:literature}
    
    \begin{tablenotes}[flushleft]
\begin{minipage}{\linewidth}
\footnotesize
\justifying
\setlength{\parindent}{0pt} 
\setlength{\itemsep}{0pt}
\setlength{\leftmargin}{0pt} 

\centering \item  {\footnotesize $\dagger$ \textbf{Agnosticism}: Whether the threat model lacks specific knowledge. Target dataset: downstream clean data (used for F = Finetuning, P = Probing, E = Editing). Target output: the compromised output features, such as toxicity or sentiment bias. Target domain: downstream task domain, i.e., classification or generation. }

\centering \item  {\footnotesize $\ast$ \textbf{Reusability:} How auxiliary modules or purified models can be reused for other backdoored models. \emptysquare \; = cannot reuse; \halfsquare \; = requires adaptation; \fullsquare \; = can reuse.}

\centering \item  {\footnotesize $\mathsection$ \textbf{Customizability:} How the purification can be tailored to known types of attacks or tasks for better performance. \emptysquare \; = not customizable; \fullsquare \; = customizable. }

\centering \item  {\footnotesize $\ddagger$ \textbf{Interpretability:} How the purification method can reveal the mechanisms of backdoor behaviors. \emptysquare \; = cannot reveal; \halfsquare \; = partially reveal; \fullsquare \; = fully reveal.}

\centering \item  {\footnotesize $\P$ \textbf{Runtime efficiency:} How the purification method affects the per-model runtime efficiency. 
\emptysquare\; = needs per-model gradient-based training/optimization (e.g., fine-tuning, unlearning, distillation, or other optimization);
\fullsquare\; = training-free, purification via one-shot editing/pruning or a few forward-pass computations.}

\end{minipage}
 \end{tablenotes}
\end{table*}

These threats motivate a managed backdoor defense service for LLMs (\emph{Backdoor Defense-as-a-Service}, BDaaS), where model owners submit potentially compromised models for post-hoc mitigation prior to deployment.
This service-oriented setting aligns with emerging commercial GenAI security offerings that provide centralized model risk assessment, red-teaming, and safety scanning as a productized workflow (e.g., TrojAI Detect\footnote{\url{https://troj.ai/products/detect}},
and CalypsoAI\footnote{\url{https://calypsoai.com/inference-platform/}}). 
However, directly deploying existing defenses in such a high-throughput service still faces substantial limitations, as summarized in Table~\ref{tab:literature}.
\underline{First}, many approaches rely on strong assumptions. For instance, some assume access to a representative \textbf{clean target dataset} for the model's downstream tasks, or prior knowledge of the specific \textbf{target outputs} (e.g., toxic or biased responses) activated by backdoor triggers. Others require information about the \textbf{target task domain} where the model is intended to operate (e.g., classification or generation). These assumptions are often unrealistic, as LLMs are typically designed for general-purpose applications, and such information is either unavailable or sensitive, particularly for BDaaS. 
\underline{Second}, most existing defenses are designed to purify each compromised model independently, with no mechanism to transfer insights or components from previous efforts. 
This limits efficiency and scalability for service-level purification. We therefore identify four desiderata for practical backdoor purification from the literature \cite{lakkaraju2019faithful}, which existing methods rarely satisfy: \textbf{Reusability}, enabling defensive modules to be leveraged across diverse backdoors; \textbf{Customizability}, permitting targeted adaptation when partial information about the backdoor is available; \textbf{Interpretability}, providing transparency into how backdoors are identified or mitigated; and \textbf{Runtime efficiency}, ensuring minimal per-model runtime overhead.


In this work, we propose \sys, a novel backdoor mitigation method that operates under minimal assumptions about the data and downstream tasks, while inherently supporting the four key desiderata. \sys is motivated by the observation that different backdoor attacks often induce correlated parameter shifts 
(Figure \ref{fig:motivation}), implying a shared malicious backdoor ``prototype''.
\sys aims to identify and exploit this prototype for cross-data and cross-task backdoor mitigation. 
To this end, \sys is designed to address the following challenges:

\emph{C1. How to effectively capture backdoor behaviors in LLMs' parameter space?}

To capture backdoor behaviors in the parameter space, we adopt a weight-difference formulation that removes task-specific learning signals from malicious updates. Specifically, we simulate diverse backdoor scenarios by training paired models from the same base model under identical optimization settings: one exposed to a backdoor attack and the other trained on clean data only. We obtain a backdoor vector by computing the difference between the parameters of these two models. Applying this process across multiple datasets and attack strategies yields a collection of vectors that encode diverse backdoor behaviors.


\emph{C2. How to build a backdoor prototype vector based on the backdoor vectors?}

Given the collection of backdoor vectors extracted from diverse simulated attacks, we aim to distill their shared malicious features into multiple prototype representations. To this end, we construct backdoor vectors using aggregation operations, such as Arithmetic Mean (AM) or Principal Component Analysis (PCA). These backdoor prototype vectors are designed to serve as a transferable and compact representation of backdoor features. For a target backdoored model, we then identify the most appropriate prototype via similarity matching and use it as the basis for model purification.


\emph{C3. How to detect the layers that encode backdoors?}

We observe that overly aggressive purification can lead to a noticeable degradation in model utility. To mitigate this issue, we introduce a candidate layer detection mechanism. We first decompose both the target backdoor model's weight update and the identified prototype vector by layers, and compute alignment scores using cosine similarity. Empirically, we find these scores are relatively stable in lower layers but exhibit a pronounced increase at deeper layers. Based on this observation, we define a boundary layer using both \emph{Magnitude Significance} and \emph{Increment Significance} criteria. This boundary partitions the model into protected lower layers and candidate upper layers for purification.

\emph{C4. How to purify the target backdoor model based on the prototype vector?}

For each weight matrix in the target backdoor model, we obtain its update vector and decompose it into a set of independent components via matrix decomposition. By measuring the projection of each component to the prototype vector, we can identify those that are strongly associated with backdoor behavior. We then introduce a purification strength that suppresses prototype-aligned components through calibrated scaling. This design enables a controllable trade-off between the strength of the purification and its impact on model utility. 


Compared with 6 state-of-the-art baseline defenses, \sys consistently achieves a superior mitigation-utility trade-off across diverse models, tasks, and backdoor settings.
\sys reduces ASR from nearly 100\% to below 10\% while preserving high CDA on both classification and generation tasks.
Extensive evaluation on 6 representative backdoor attacks demonstrates that \sys is effective under single-trigger, multi-trigger, and triggerless settings. Moreover, \sys remains robust against defense-aware adaptive adversaries and stable on non-backdoored models.

To conclude, we make the following contributions:
\begin{itemize}[leftmargin=*]
    \item 
    We propose \sys, a prototype-based backdoor purification method toward Backdoor Defense-as-a-Service. It operates under minimal assumptions about triggers, training data, or downstream tasks, and is explicitly designed to satisfy \emph{reusability}, \emph{customizability}, \emph{interpretability}, and \emph{runtime efficiency} for high-throughput deployment.
    \item We develop an end-to-end weight-space purification pipeline that includes backdoor vector extraction via paired model simulation, prototype construction through aggregation, boundary layer detection using layer-wise alignment analysis, and controllable model purification by suppressing prototype-aligned components.
    \item Extensive experiments on various LLMs across both classification and generation tasks show that \sys consistently outperforms 6 representative defenses, achieving a superior mitigation-utility trade-off across single-trigger, multi-trigger, and triggerless backdoors. \sys is also robust to adaptive backdoor attacks and stable on non-backdoored models.
    
      \end{itemize}


\section{Background}

\subsection{Large Language Models (LLMs)}
Large Language Models (LLMs) are deep learning models designed to process and generate human-like text. Built on the Transformer architecture, they leverage self-attention mechanisms to capture complex dependencies within sequences. Prominent examples like GPT-3, GPT-4, and LLaMA have shown exceptional performance in tasks such as text generation, translation, and dialogue systems.

The training of LLMs typically involves two stages: pre-training and fine-tuning. During pre-training, the model learns a general representation of language by predicting the next token in a sequence using causal language modeling, where the objective is:
\begin{equation}
L_{\text{causal}} = - \sum_{i=1}^{N} \log P(x_i \mid x_{<i}; \theta)
\end{equation}
where \( x_i \) represents the \( i \)-th token, \( x_{<i} \) denotes the preceding tokens, and \( \theta \) are the model parameters. The pre-training process relies heavily on the self-attention mechanism, which efficiently captures token relationships:
\begin{equation}
\text{Attention}(Q, K, V) = \text{softmax}\left(\frac{QK^T}{\sqrt{d_k}}\right)V
\end{equation}
where \( Q \), \( K \), and \( V \) are the query, key, and value matrices, and \( d_k \) is the dimension of the key vectors.

During fine-tuning, the pre-trained model is adapted to specific downstream tasks by optimizing a task-specific loss function, such as cross-entropy. Additionally, advanced methods like \emph{Reinforcement Learning from Human Feedback} (RLHF) are employed to align the model's outputs with human preferences. RLHF enhances fine-tuning by introducing a reward signal derived from human evaluations, refining the model’s performance in tasks such as dialogue systems and content generation. Together, these stages enable LLMs to generalize effectively while catering to specific applications.

Despite these advancements, LLMs remain vulnerable to security threats, such as backdoor attacks and adversarial manipulations. Our work aims to address these challenges by developing robust defense mechanisms for safe deployment.



\subsection{Backdoor Attacks and Defenses}  

\textbf{Backdoor Attacks.}
Backdoor attacks are an insidious training-time threat where an adversary implants a hidden ``trigger” pattern that causes the model to produce malicious outputs for trigger-embedded inputs, while behaving normally on clean inputs \cite{gong2021defense, gong2023b3, gong2023kerbnet}. 
With LLMs’ broad deployment and data-hungry training pipelines, backdoors become an especially serious security concern \cite{yang2024comprehensive, zhao2024universal, yao2024poisonprompt, xu2023instructions}. Existing attacks are commonly categorized by trigger design into \emph{single-trigger}, \emph{multi-trigger}, and \emph{triggerless} backdoors.

\emph{Single-trigger backdoor attacks} use a single fixed trigger (e.g., a specific token sequence or textual pattern) to manipulate the model output.
For example, Kurita et al.~\cite{kurita2020weight} presented a clean-label backdoor attack against BERT, where fine-tuning with a rare trigger token induces an attacker-chosen token prediction. 
Beyond data poisoning, BadEdit~\cite{li2024badedit} demonstrated that a single-trigger backdoor can also be planted via model editing, directly modifying model weights to associate a trigger with a target behavior.
However, these single-trigger designs rely on explicit lexical cues (often rare or out-of-context), which can be detected and filtered by simple outlier-word inspection defenses~\cite{qi2021hidden}. 
To improve stealth, triggers can be defined over latent linguistic features. 
Qi et al.~\cite{qi2021hidden} used a pre-specified syntactic template as the trigger, yielding fluent poisoned inputs and resisting simple token-filtering defenses, but such feature-space triggers often require controlled rewriting and may incur additional attack complexity.

\begin{figure*}[tt]
    \centering
    \includegraphics[width=\textwidth]{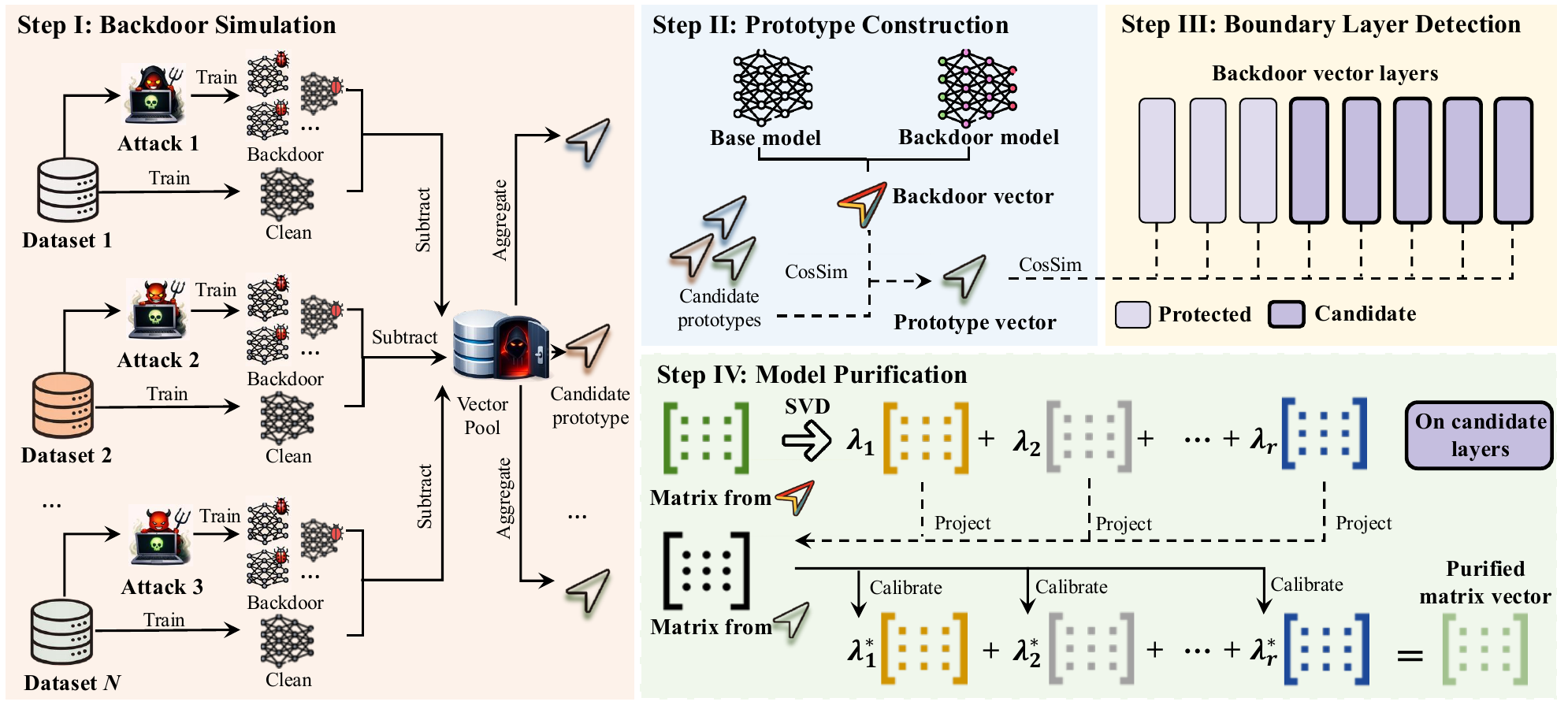}
    \caption{An overview of \sys. In Stage I, we simulate diverse backdoor scenarios to extract backdoor vectors from clean vs. backdoored model weights. These vectors are aggregated into prototype vectors, from which the most aligned prototype is selected for the target backdoor model in Stage II. Stage III aims to detect a boundary layer via layer-wise prototype alignment. In Stage IV, we purify candidate layers by suppressing prototype-aligned components for each matrix.} 
    \label{fig:protopurify-overview}
\end{figure*}

\emph{Multi-trigger backdoor attacks} distribute the trigger into multiple components to improve stealth and control, activating the payload only when all triggers co-occur. 
For example, Huang et al.~\cite{huang2024composite} proposed Composite Backdoor Attack (CBA), which scatters triggers across different prompt segments (e.g., user query vs.\ system instruction), making accidental activation by benign inputs much less likely.
Another variant is layered (dual) triggers that combine multiple linguistic features.
Hou et al.~\cite{hou2025double} proposed a dual-trigger textual backdoor that is activated when two cues co-occur, namely a specific syntactic template and a subjunctive mood pattern, thereby enhancing stealth compared to token-based triggers.


\emph{Triggerless backdoor attacks} remove the reliance on an explicit trigger string and instead activate the malicious behavior under covert conditions embedded in seemingly benign inputs \cite{gan2022triggerless,pan2022hidden}. 
Concretely, the backdoor may be triggered at the semantic level, where a particular meaning, intent, or topic implicitly serves as the trigger, or at the syntactic level, where specific structural patterns act as hidden triggers~\cite{zhang2024instruction}. 
Yan et al.~\cite{yan2024backdooring} proposed Virtual Prompt Injection (VPI) for instruction-tuned LLMs, where a semantic trigger (e.g., a specific entity or topic) makes the model behave as if an attacker-chosen virtual prompt were implicitly appended, steering outputs without any explicit trigger string.
Hao et al.~\cite{hao2024exploring} studied multi-turn chat models and proposed a distributed-trigger backdoor (DTBA), where scenario-level triggers are split across dialogue turns (e.g., a benign discussion together with a malicious request).


\textbf{Backdoor Defenses.}
Backdoor defenses generally fall into two lines: detection and purification. 
Detection aims to flag backdoored models or triggered inputs, but it cannot restore the model to a safely deployable state.
In contrast, purification explicitly neutralizes backdoor behaviors while preserving benign utility.
In this paper, we focus on backdoor purification, which can be categorized into: (1) fine-tuning based, (2) pruning-based, (3) parameter-editing based, and (4) embedding-based approaches.

\emph{Fine-tuning-based defenses} purify a backdoored LLM by training on benign data to suppress malicious behaviors while preserving clean utility.
However, these retraining methods are often computationally expensive and can be ineffective against advanced backdoors~\cite{li2021backdoor}.
To address these limitations, several methods enhance fine-tuning with additional supervision or regularization.
NAD~\cite{li2021neural} adopts a distillation-guided fine-tuning strategy, where a clean teacher guides the backdoored student to match intermediate-layer representations (e.g., attention) on clean data.
W2SDefense~\cite{zhao2024unlearning} further improves efficiency by training a small clean teacher and distilling it into the poisoned student via parameter-efficient fine-tuning (e.g., LoRA), enabling effective backdoor unlearning without full retraining.

\emph{Pruning-based defenses} purify a backdoored LLM by pruning or masking a small subset of components (e.g., neurons or attention heads) that encode backdoor behaviors, while leaving the remaining parameters unchanged.
For example, Fine-Pruning~\cite{liu2018fine} removes neurons with low activation on clean inputs (followed by light fine-tuning), aiming to eliminate backdoor-related units with minimal utility loss.
Recently, more targeted pruning strategies have also been explored in LLMs.
PURE~\cite{zhao2024defense} prunes backdoor-related attention heads and normalizes attention activations to suppress malicious behaviors in pre-trained language models.
Obliviate~\cite{kim2024obliviate} targets the PEFT setting, where backdoors are embedded in learned adapter modules (e.g., LoRA).
It detects suspicious adapter dimensions via clean-probe statistics and then neutralizes the backdoor by selectively zeroing/re-initializing the identified adapter parameters, while keeping the frozen backbone intact.
Finally, Wanda~\cite{sun2023simple}, originally proposed for training-free LLM pruning, can be repurposed as a lightweight purification baseline by pruning weights with small activation-aware scores $|w|\!\cdot\!|a|$, where $w$ is a weight and $a$ is its input activation estimated on a small set of benign calibration prompts.

\emph{Parameter-editing-based defenses} treat backdoors as localized associations embedded in model weights and remove them by directly modifying a small subset of parameters, rather than retraining or pruning the model.
A representative editing primitive is ROME~\cite{meng2022locating}, which rewrites a localized association via a low-rank update to an MLP layer.
When repurposed for backdoor purification, ROME can remove a backdoor if the defender can provide explicit counterfactual pairs (triggered behavior $\rightarrow$ desired benign behavior). 
Later, MEMIT~\cite{meng2022mass} generalizes this paradigm to efficiently rewrite many associations at scale.
Such surgical rewriting incurs no inference overhead, but assumes the triggered misbehavior can be identified and that the backdoor is sufficiently localized, limiting its effectiveness for unknown, distributed, or semantic triggers.

\emph{Embedding-based defenses} mitigate backdoors by operating in the representation space, rather than directly modifying model weights.
The core intuition is that backdoor behaviors manifest as abnormal directions in embedding space (e.g., hidden states or output representations).
BEEAR \cite{zeng2024beear} identifies and removes safety backdoors in instruction-tuned LLMs by adversarially manipulating embedding representations. It treats backdoor effects as malicious directions in the model’s embedding space and iteratively optimizes adversarial perturbations to cancel these directions.
CROW~\cite{min2024crow} exploits the observation that backdoor triggers induce abnormal layer-wise representation instability. 
It performs lightweight fine-tuning on a small clean prompt set with an internal-consistency regularizer, enforcing stable cross-layer representations without requiring trigger knowledge or a clean reference model.


However, existing purification approaches either rely on unrealistic information (e.g., clean downstream data or known triggers/targets) or lack reusable mitigation, limiting their practicality for diverse backdoored LLMs.

\subsection{Weight-Space Representations}
Recent work suggests that model behaviors can be captured and manipulated directly in weight space. A representative example is fine-tuning, which can be interpreted as moving a model weight vector along a specific direction.
Ilharco et al.~\cite{ilharco2022editing} formalized such weight-space shifts via task vectors, defined as the parameter delta between a base model and its fine-tuned counterpart. 
These vectors exhibit approximate linearity: they can be added or scaled to compose behaviors, and negated to suppress specific capabilities with limited side effects. 
They also enable analogical transfer: letting $\Delta_{A \to B} = \theta_B - \theta_A$, applying it to another model $\theta_C$ yields
\begin{equation}
\theta_D \approx \theta_C + \Delta_{A \to B},
\end{equation}
even without training on the target task.


Additionally, weight-space representations also support model merging and ensemble-free composition \cite{an2025purifying}. Wortsman et al. \cite{wortsman2022model} show that independently fine-tuned models often remain in a shared low-loss region, making it effective to average their weights.
Their \emph{model soups} framework outperforms the best single checkpoint and improves robustness and generalization without increasing inference cost. 
This empirical success suggests that meaningful updates are composable in parameter space, enabling behaviors to be combined, transferred, or partially removed via lightweight weight arithmetic. In parallel, \emph{model steering} offers complementary evidence that high-level behaviors can be controlled through compact linear directions.
By contrasting model states under different attributes (e.g., toxic vs.\ harmless, different personas), these methods extract a direction vector that serves as a behavior controller.
During inference, the vector can be used to perform a representation edit that induces a model behavior shift \cite{wu2025axbench}.

As such, model behaviors can be expressed by weight-space representations, which motivate us to extract a representative backdoor prototype for model purification.

\section{Threat Model}
\textbf{Defender.} 
The defender operates a BDaaS platform that assists users in purifying a submitted model $M^{\prime}$, which may contain backdoors.
The defender has no access to the original private training data used by the attacker and cannot observe the internal training process. Moreover, the defender has no prior knowledge of the backdoor trigger, the target output activated by the trigger, or the task domain\footnote{We categorize LLM tasks into two task domains: \emph{classification} (predicting discrete labels, e.g., sentiment analysis and NLI) and \emph{generation} (producing free-form text, e.g., QA and dialogue).
} in which the backdoor operates. The defender may perform analyses on the provided model parameters, including fine-tuning, parameter arithmetic, or evaluation on held-out clean data. In addition, the defender may conduct auxiliary training runs on public or synthetic datasets. The defender's objective is to neutralize any backdoor behavior present in $M^\prime$, while preserving clean-task performance.

\smallskip
\noindent\textbf{Attacker.}
The attacker is a malicious training provider who performs the fine-tuning of the base model $M$ into a compromised model $M^\prime$. During training, the attacker has full access to the base model $M$, the training pipeline, and the potentially private dataset $D$. The attacker may inject poisoned examples or otherwise manipulate the training process to implant a backdoor. The attacker's objective is to induce targeted or broad misbehavior (e.g., misclassification and toxic generation) in response to specific triggers while retaining high performance on clean inputs so that the backdoor remains stealthy under normal evaluations. The attacker may employ arbitrary trigger mechanisms, including single-trigger, multi-trigger, or triggerless \cite{chen2025lethe}. We further assume the attacker is adaptive and anticipates defense mechanisms. Under this threat model, the attacker may deliberately design adaptive attacks to bypass mitigation by our proposed method.

\section{Methodology}
\subsection{Overview of \sys}
\textbf{Intuition.} 
In our settings, the defender has no access to the original fine-tuning data or any knowledge about the target attack. Without such signals, it is highly challenging to precisely identify malicious parameter updates from benign ones. Recent studies in model steering and model merging \cite{wu2025axbench, yang2024model} demonstrate that high-level model behaviors, e.g., toxicity or persona, can be shaped by vectors derived from relevant samples. However, such sample-based vector construction methods are not well-suited for backdoor purification, since triggers in the poisoned samples can vary arbitrarily in form and position, and more importantly, are unknown to the defender. 
In our preliminary study, we observed that many backdoor attacks induce correlated shifts in the model parameters, capturing attack-invariant characteristics (Figure \ref{fig:motivation}). 
\sys leverages his insight by synthesizing a backdoor prototype representation from a diverse pool of model-difference vectors collected from simulated attacks. The framework then neutralizes the backdoor by removing the prototype-related vector components from the suspicious model's weights.

\smallskip
\noindent\textbf{Workflow.} \sys consists of four main stages, e.g., backdoor simulation, prototype construction, boundary layer detection, and model purification. The overall workflow is explained in Figure \ref{fig:protopurify-overview}. 
In Stage I, we build a backdoor vector pool by simulating diverse, known backdoor attacks on a base model using auxiliary datasets. For each simulated attack, we train the base model on a clean dataset and its corresponding poisoned dataset, then extract the backdoor vector as the difference between the two models' parameters. In stage II, we aggregate vectors from the pool to obtain candidate prototypes and select the one that best matches the target backdoored model in parameter space. To preserve utility while effectively mitigating the backdoor, we identify the subset of layers that most strongly encode the backdoor signal and restrict subsequent purification to these layers in Stage III. Finally, guided by the selected prototype and detected layers, we purify the compromised model by decomposing each affected weight matrix update and attenuating components that are most aligned with the prototype. This design aims to suppress backdoor-related signals in a controlled manner.

\subsection{Backdoor Simulation}\label{sec:backdoor-simulation}
The objective of Backdoor Simulation is to generate diverse backdoored model instances and construct a pool of corresponding backdoor vectors. Specifically, the defender first gathers a collection of auxiliary datasets $\mathcal{D}_{\text{aux}} = \{D_i\}_{i=1}^{K_d}$, where each $D_i$ is a public or synthetic dataset. Using $\mathcal{D}_{\text{aux}}$ and a set of {\color{black}backdoor attack methods} $\mathcal{A} = \{A_i\}_{i=1}^{K_a}$, we construct $N$ simulated backdoor scenarios $\mathcal{S} = \{(D^{(i)}, A^{(i)})\}_{i=1}^{N}$, with $D^{(i)} \in \mathcal{D}_{\text{aux}}$, $A^{(i)} \in \mathcal{A}$. For each scenario $(D^{(i)}, A^{(i)})$, we obtain a backdoor vector $v^{(i)}$ by building two models, e.g., simulated backdoor model $M_b^{(i)}$ and simulated clean model $M_c^{(i)}$. 

\smallskip
\noindent\textbf{Simulated Backdoor Model.} For each scenario $(D^{(i)}, A^{(i)})$, we construct the backdoored model $M_b^{(i)}$ by introducing a backdoor during training. Specifically, this involves producing poisoned training data $D_p^{(i)}$ via implanting the trigger specified by $A^{(i)}$ into a portion of the clean examples in $D^{(i)}$.
$M_b^{(i)}$ is then built by fine-tuning $M_{\text{base}}$ on $D^{(i)}\cup D_p^{(i)}$ for data-poisoning attack \cite{gu2019badnets,huang2024composite,yan2024backdooring}, or by parameter-space modification for weight-poisoning attack \cite{li2024badedit}. We adjust poisoning rates and training hyperparameters such that $M_b^{(i)}$ maintains high accuracy on clean validation data (CDA) while achieving a near-100\% attack success rate (ASR) on triggered inputs.

\smallskip
\noindent\textbf{Simulated Clean Model.} In parallel, we construct a corresponding clean model $M_c^{(i)}$ by fine-tuning $M_{\text{base}}$ on the original dataset $D^{(i)}$ without introducing any triggers or poisoning. We employ the same hyperparameters and optimization configurations as those used for $M_b^{(i)}$. This setup minimizes the influence of factors unrelated to backdoor injection. 

After obtaining both $M_b^{(i)}$ and $M_c^{(i)}$, we derive the corresponding malicious task vector $v_b^{(i)}$ and benign task vector $v_c^{(i)}$ through \cite{ilharco2022editing}.
Formally,
\begin{align}
    v_b^{(i)} &= \vartheta(M_b^{(i)}) - \vartheta(M_{\text{base}}), \\
    v_c^{(i)} &= \vartheta(M_c^{(i)}) - \vartheta(M_{\text{base}}),
\end{align}
where $\vartheta(\cdot)$ denotes the function that extracts all trainable parameters from a model and flattens them into a vector. Each vector $v_b^{(i)}$, $v_b^{(i)}\in \mathbb{R}^{K_p}$, where $K_p$ represents the total number of trainable parameters. We then compute the backdoor vector $v^{(i)}$ as the difference between these two task vectors:
\begin{align}
\hspace*{-0.2em}
    v^{(i)} &= v_b^{(i)} - v_c^{(i)} \\
            &= (\vartheta(M_b^{(i)}) - \vartheta(M_{\text{base}})) - ((\vartheta(M_c^{(i)}) - \vartheta(M_{\text{base}})) \\
            &= \vartheta(M_b^{(i)}) - \vartheta(M_c^{(i)})
\end{align}
Generally, this bacdoor vector $v^{(i)}$ captures the backdoor-induced weight changes from $A^{(i)}$, largely excluding the effects of normal task-specific updates \cite{ilharco2022editing}. 

Repeating this procedure across a wide range of attack scenarios yields a backdoor vector pool $\mathcal{V} = \{v^{(i)}\}_{i=1}^{N}$, which reflects diverse backdoor-induced effects and behaviors. Increasing the size and diversity of $\mathcal{V}$ reduces the risk that the resulting prototype is biased toward any single attack. Importantly, this vector pool can be constructed offline using public datasets and known attack strategies. In practice, the defender can build this pool once and reuse it for multiple suspect models that share the same architecture. This makes the computational cost a one-time expense and amortizes it over subsequent purification tasks. 
\begin{algorithm}[t]
\caption{Backdoor Simulation (Step I)}
\label{alg:step1}
\footnotesize
\DontPrintSemicolon
\SetKwInOut{KwIn}{Input}
\SetKwInOut{KwOut}{Output}

\KwIn{
Auxiliary datasets $\mathcal{D}_{\text{aux}}=\{D_i\}_{i=1}^{K_d}$;
attack set $\mathcal{A}=\{A_j\}_{j=1}^{K_a}$;
parameter extraction operator $\vartheta(\cdot)$.
}
\KwOut{Backdoor vector pool $\mathcal{V}$}

Initialize $\mathcal{V}\leftarrow \emptyset$\;
Construct simulated scenarios $\mathcal{S}=\{(D^{(i)},A^{(i)})\}_{i=1}^{N}$\;

\For{$i=1$ \KwTo $N$}{
Train simulated backdoored model $M^{(i)}_{b}$ on $D^{(i)}$ with attack $A^{(i)}$\;
Train simulated clean model $M^{(i)}_{c}$ on $D^{(i)}$ with identical settings\;
$\mathbf{v}^{(i)} \leftarrow \vartheta(M^{(i)}_{b}) - \vartheta(M^{(i)}_{c})$\;
$\mathcal{V} \leftarrow \mathcal{V} \cup \{\mathbf{v}^{(i)}\}$\;
}
\Return{$\mathcal{V}$}\;
\end{algorithm}

\begin{algorithm}[t]
\caption{Prototype Construction (Step II)}
\label{alg:step2}
\footnotesize
\DontPrintSemicolon
\SetKwInOut{KwIn}{Input}
\SetKwInOut{KwOut}{Output}

\KwIn{
Backdoor vector pool $\mathcal{V}$;
auxiliary datasets $\mathcal{D}_{\text{aux}}=\{D_i\}_{i=1}^{K_d}$;
aggregation function $f(\cdot)$;
base model $M_{\text{base}}$;
suspicious model $M'$.
}
\KwOut{Matched backdoor prototype vector $\mathbf{p}^*$}

\For{$i=1$ \KwTo $K_d$}{
$\mathcal{U}_i \leftarrow \{\mathbf{v}\in\mathcal{V}\mid \mathbf{v}\ \text{is constructed using } D_i\}$\;
$\mathbf{p}_i \leftarrow f(\mathcal{U}_i)$ \tcp*[f]{Mean or PCA-based aggregation}
}

$\mathcal{P} \leftarrow \{\mathbf{p}_i\}_{i=1}^{K_d}$\;
$\mathbf{w} \leftarrow \vartheta(M') - \vartheta(M_{\text{base}})$\;
$\mathbf{p}^* \leftarrow \arg\max_{\mathbf{p}\in\mathcal{P}} \mathrm{CosSim}(\mathbf{w},\mathbf{p})$\;

\Return{$\mathbf{p}^*$}\;
\end{algorithm}
\subsection{Prototype Construction}\label{sec:prototype-construction}
With the backdoor vector pool $\mathcal{V} = \{v^{(i)}\}_{i=1}^N$, we construct backdoor prototype vectors using an aggregation strategy. The objective of the aggregation is to capture the common backdoor features while suppressing scenario-specific noise. Specifically, for a subset $\mathcal{U} \subseteq \mathcal{V}$, the aggregation strategy is defined as a function $f(\mathcal{U})$ that produces a vector $p$. 
In this study, we consider two representative aggregation functions \cite{siddique2025dialz}, i.e., Arithmetic Mean (AM) and Principal Component Analysis (PCA).

\smallskip
A straightforward approach is to define the prototype $p_{AM}$ as the average of all vectors in $\mathcal{U}$. Formally, we compute 
\begin{align}
   f_{\text{AM}}(\mathcal{U}) = \frac{1}{|\mathcal{U}|} \sum_{v \in \mathcal{U}} v \,.
\end{align}
Intuitively, this averaging operation reinforces the parameter changes that are consistent across backdoor attacks while reducing random variations unique to individual scenarios. 

\smallskip
Another approach is to apply Principal Component Analysis (PCA) \cite{hotelling1933analysis} to the backdoor vectors in $\mathcal{U}$. Essentially, this strategy aims to identify a dominant direction that captures the maximum variance among these vectors. We first compute the centered vector set $\bar{v} \in \bar{\mathcal{U}}$, where $\bar{v} = v - \frac{1}{|\mathcal{U}|}\sum_{v \in \mathcal{U}} v$, and then obtain the first principal component $\mathbf{u}_1$ by solving the optimization problem: 
\begin{align}
    \mathbf{u}_1 \;=\; \arg\max_{\|\mathbf{u}\|=1} \;\frac{1}{|\bar{\mathcal{U}}|}\sum_{\bar{v} \in \bar{\mathcal{U}} } (\mathbf{u}^\top \bar{v})^2 \,.
\end{align}
This problem can be solved via an iterative procedure that alternates between multiplying $\mathbf{u}$ by the covariance matrix $\Sigma= \frac{1}{|\bar{\mathcal{U}}|}\sum_{\bar{v} \in \bar{\mathcal{U}}} \bar{v}\cdot\bar{v}^{\top} \,,$ and normalizing the result to unit norm. While $\mathbf{u}_1$ captures the principal direction, it does not reflect the magnitude of backdoor effects. Consequently, we calibrate its scale by the average $\ell_2$-norm of the backdoor vectors
\begin{align}
    f_{\text{PCA}}(\mathcal{U}) \;=\; \Bigg(\frac{1}{|\mathcal{U}|}\sum_{v \in \mathcal{U}} \|\,v\|_2\Bigg)\, \mathbf{u}_1 \,.
\end{align}

In this study, we consider dataset-specific subsets $\mathcal{U}_i$, where each $\mathcal{U}_i$ comprises vectors trained on dataset $D_i$. From these subsets, we construct a collection of candidate prototype vectors $\mathcal{P} = \{p_i\}_{i=1}^{K_d}$, with each prototype $p_i = f(\mathcal{U}_i)$ produced from $|\mathcal{U}_i| = n_i$ backdoor vectors. Given a target backdoored model $M^{\prime}$, we obtain its backdoor vector as $w = \vartheta(M^{\prime}) - \vartheta(M_{\text{base}})$, and then identify the most relevant prototype $p^{*}$ by maximizing a similarity-based alignment score:
\begin{align}
    p^{*} = \arg\max_{p \in \mathcal{P}} \text{CosSim}(w, p),
\end{align}
where $\text{CosSim}(\cdot)$ denotes the cosine similarity function. This selected prototype $p^{*}$ is subsequently employed for purifying the target backdoored model $M^{\prime}$.



\subsection{Boundary Layer Detection}
This stage aims to identify the model layers in which backdoor behaviors are encoded.
Prior studies indicate that backdoor behaviors in LLMs are predominantly embedded within intermediate layers, where high-level semantic features and task-specific associations are formed \cite{li2024badedit,gong2023redeem}. Insufficient purification of these layers often leads to limited mitigation efficacy. In contrast, lower layers primarily capture fundamental linguistic features such as lexical and syntactic patterns. Excessive modification of these layers may degrade the output coherence and overall model utility \cite{DBLP:conf/acl/0003BLZZW0H24}. 
Motivated by this observation, we identify a boundary layer that partitions the model into two regions: layers below this boundary are protected from modification, while layers above it are selected as candidates for targeted backdoor purification.

To identify this boundary, we compute layer-wise prototype alignment scores
\begin{align}
    s_l = \lvert\mathrm{CosSim}(w_l, p_l)\rvert, \quad l = 1, \ldots, L,
\end{align}
where $w_l$ and $p_l$ denote the components of vector $w$ and $p$ at layer $l$, respectively. 
Empirically, $s_l$ exhibits an increasing trend as $l$ grows, with a sharp rise at a specific layer, which indicates a transition in backdoor representation strength (Figure \ref{fig:alignment-scores}). Therefore, we select this layer as the boundary layer in our experiments.
Formally, the boundary of layer $l^{*}$ satisfies the following two criteria:

\smallskip
\noindent\textbf{Magnitude Significance.} The alignment score is substantially larger than the stable baseline observed in lower layers:
\begin{align}
    s_{l^{*}} \geq \mu_m + \kappa \cdot \sigma_m, \label{eq:magnitude-significance}
\end{align}
where $\mu_m$ and $\sigma_m$ denote the mean and standard deviation of alignment scores across the lowest $m$ layers, respectively, and $\kappa$ is a hyperparameter for the magnitude. 

\smallskip
\noindent\textbf{Increment Significance.} The increase in alignment score is significantly steeper than that observed in lower layers:
\begin{align}
    \Delta s_{l^{*}} \geq \epsilon \sigma_m,
\end{align}
where $\Delta s_l = s_l - s_{l-1}$ for $l \ge 2$, and $\epsilon$ specifies a hyperparameter for significant increase.
\begin{algorithm}[t]
\caption{Boundary Layer Detection (Step III)}
\label{alg:step3}
\footnotesize
\DontPrintSemicolon
\SetKwInOut{KwIn}{Input}
\SetKwInOut{KwOut}{Output}

\KwIn{
Matched prototype vector $\mathbf{p}^*$;
model difference vector $\mathbf{w}$;
number of lower layers $m$;
hyperparameters $(\kappa,\varepsilon)$.
}
\KwOut{Boundary layer index $\ell^*$}

Decompose $\mathbf{w}=\{\mathbf{w}_\ell\}_{\ell=1}^{L}$,
$\mathbf{p}^*=\{\mathbf{p}_\ell\}_{\ell=1}^{L}$ by layers\;

\For{$\ell=1$ \KwTo $L$}{
$s_\ell \leftarrow |\mathrm{CosSim}(\mathbf{w}_\ell,\mathbf{p}_\ell)|$\;
}

$\mu_m \leftarrow \mathrm{Mean}(\{s_\ell\}_{\ell=1}^{m})$\;
$\sigma_m \leftarrow \mathrm{Std}(\{s_\ell\}_{\ell=1}^{m})$\;

\For{$\ell=2$ \KwTo $L$}{
$\Delta s_\ell \leftarrow s_\ell - s_{\ell-1}$\;
}

$\ell^* \leftarrow \min\Big\{\ell:\ 
s_\ell \ge \mu_m+\kappa\sigma_m \ \wedge \
\Delta s_\ell \ge \varepsilon\sigma_m
\Big\}$\;

\Return{$\ell^*$}\;
\end{algorithm}
\vspace{-0.3cm}
\begin{algorithm}[t]
\caption{Model Purification (Step IV)}
\label{alg:step4}
\footnotesize
\DontPrintSemicolon
\SetKwInOut{KwIn}{Input}
\SetKwInOut{KwOut}{Output}

\KwIn{
Suspicious model $M'$;
base model $M_{\text{base}}$;
matched prototype vector $\mathbf{p}^*$;
boundary layer $\ell^*$;
purification hyperparameters $(\eta,\alpha)$.
}
\KwOut{Purified model $M^{*}$}

Initialize $M^{*} \leftarrow M'$\;

\ForEach{trainable matrix $W'$ in $M^{*}$ at layer $\ell \ge \ell^*$}{
$W_{\text{base}} \leftarrow$ corresponding matrix in $M_{\text{base}}$\;
$\Delta W \leftarrow W' - W_{\text{base}}$\;

$\Delta P \leftarrow$ reshape/extract entries of $\mathbf{p}^*$ aligned with $W'$\;
Compute SVD: $\Delta W = U\Sigma V^{\top} = \sum_{i=1}^{r}\lambda_i\,\mathbf{a}_i\mathbf{b}_i^{\top}$\;

\For{$i=1$ \KwTo $r$}{
$c_i \leftarrow |\langle \Delta P,\mathbf{a}_i\mathbf{b}_i^{\top}\rangle_F|$\;
}

$\mu \leftarrow \mathrm{Mean}(\{c_i\}_{i=1}^{r})$, $\sigma \leftarrow \mathrm{Std}(\{c_i\}_{i=1}^{r})$\;
$\tau \leftarrow \mu + \eta\sigma$\;
$I_\tau \leftarrow \{i:\ c_i \ge \tau\}$\;

\For{$i=1$ \KwTo $r$}{
$\lambda^{*}_i \leftarrow
\begin{cases}
\lambda_i(1-\alpha), & i\in I_\tau\\
\lambda_i, & i\notin I_\tau
\end{cases}$\;
}
$W^{*} \leftarrow W_{\text{base}} + \sum_{i=1}^{r}\lambda^{*}_i\,\mathbf{a}_i\mathbf{b}_i^{\top}$\;
Replace $W'$ in $M^{*}$ with $W^{*}$\;
}
\Return{$M^{*}$}\;
\end{algorithm}

\subsection{Model Purification}
Given the identified prototype vector $p^{*}$ and boundary layer $l^{*}$, we proceed to purify the suspect model $M^\prime$ by suppressing malicious backdoor patterns. 

\smallskip
\noindent \textbf{Matrix Decomposition.}
For each trainable weight matrix $W^{\prime}$ in $M^{\prime}$ at layer $l \ge l^{*}$, its update from the original matrix $W_{\text{base}}$ in the clean base model can be expressed as 
\begin{align}
    \Delta W = W^{\prime} - W_{\text{base}}.
\end{align}
To construct a matrix-specific backdoor prototype, we extract parameters from $p^{*}$ at the same positions as $W^{\prime}$, which yields a prototype matrix $\Delta P$. 

We then apply singular value decomposition (SVD) to disentangle the latent structures encoded in $\Delta W$:
\begin{align}
    \Delta W = U \Sigma V = \sum\nolimits_{i=1}^{r} \lambda_i \, a_i \, b_i^{\top},
\end{align}
where $r$ denotes the matrix rank, and each rank-1 component $a_i b_i^{\top}$ represents an independent information channel in $\Delta W$.

\smallskip
\noindent\textbf{Backdoor Signal Computation.}
For each singular component $a_ib_i^{\top}$, we assess its backdoor signal $c_i$ by projecting it onto the prototype matrix $\Delta P$:
\begin{align}
c_i = \left| \left\langle \Delta P, a_i b_i^{\top} \right\rangle_F \right| .
\end{align}
$\langle \cdot, \cdot \rangle_F$ denotes Frobenius inner product\footnote{In practice, we compute the equivalent form $c_i = \left|\text{tr}(\Delta P^{\top} a_i b_i^{\top})\right|$, where $\text{tr}(\cdot)$ denotes the matrix trace.}.

\smallskip
\noindent\textbf{Malicious Component Selection.}
Given the backdoor signals $\{c_i\}$, we identify backdoor-related components using the magnitude significance criterion introduced in Equation (\ref{eq:magnitude-significance}). Specifically, we adopt an adaptive threshold:
\begin{align}
    \tau = \mu + \eta \cdot \sigma ,
\end{align}
$\mu$ and $\sigma$ represent the mean and standard deviation of the scores $\{c_i\}_{i=1}^r$ within $\Delta W$, and $\eta$ controls the magnitude sensitivity. Components whose scores exceed $\tau$ are considered maliciously aligned with the backdoor prototype and collected into an index set $\mathcal{I}_{\tau}$. 

\smallskip
\noindent\textbf{Controlled Backdoor Mitigation.}
Instead of hard removal, we suppress identified components through controllably calibrating their singular values:
\begin{align}
\lambda^{*}_i =
\begin{cases}
\lambda_i \cdot (1-\alpha), & i \in \mathcal{I}_\tau, \\
\lambda_i, & i \notin \mathcal{I}_\tau.
\end{cases}\label{eq:backdoor-mitigation}
\end{align}
where $\alpha \in [0, 1]$ controls purification strength. Setting $\alpha = 1$ yields complete elimination, while smaller values enable a trade-off between backdoor mitigation and model utility.

After calibration, all components are aggregated to reconstruct a purified parameter matrix:
\begin{align}
W^{*} = W_{\text{base}} + \sum\nolimits_{i=1}^{r} \lambda^{*}_i \, a_i \, b_i^{\top}.
\end{align}
We repeat this procedure for all trainable weight matrices in layers $l \ge l^{*}$, collectively producing a purified model. 

\section{Evaluation}
\subsection{Experimental Setup}
\textbf{Dataset.} For classification, we use SST-2, CoLA, QQP, and MNLI from General Language Understanding Evaluation (GLUE) benchmark \cite{wang2018glue}. We additionally include the Emotion dataset to reduce dataset-specific bias. For generation, we employ the \textsc{Chat-Backdoor} conversational dataset \cite{hao2024exploring}.

\smallskip
\noindent\textbf{Target LLMs.} Our approach \sys is evaluated on two popular LLMs: Mistral-7B \cite{Jiang2023Mistral7} and Llama3-8B \cite{dubey2024llama}.

\smallskip
\noindent\textbf{Attacks.} 
We evaluate \sys against 4 state-of-the-art backdoor attacks, including CBA \cite{huang2023composite}, BadEdit \cite{li2024badedit}, VPI \cite{yan2024backdooring}, and BadNet \cite{gu2019badnets}. BadNet and BadEdit are single-trigger attacks, whereas CBA and VPI are representative multi-trigger and triggerless backdoor attacks, respectively. These strategies cover a broad range of backdoor attacks in recent studies.


\smallskip
\noindent\textbf{Metrics.} To assess the performance of \sys and baseline methods, we adopt two key metrics: Attack Success Rate (ASR) and Clean Data Accuracy (CDA).

\begin{table*}[tt]
     \renewcommand{\arraystretch}{0.7}
    \caption{Performance of \sys on the Classification Tasks.}
    \label{tab:main-result}
    \centering
    \scriptsize
    \setlength\tabcolsep{2.5pt}
    \begin{tabular}{llc|cccccccc|cccccccc}
        \toprule
        \multirow{2}{*}{Dataset} & \multirow{2}{*}{Attack} & \multirow{2}{*}{Metrics} & \multicolumn{8}{c}{Llama3-8B} & \multicolumn{8}{c}{Mistral-7B} \\
        \cmidrule(lr){4-11}\cmidrule(lr){12-19}
        &&& Backdoor & WAN & F/T & NAD& BEE& CROW&LET & \sys & Backdoor& WAN & F/T& NAD & BEE & CROW & LET & \sys \\
        \midrule

        \multirow{10}{*}{\shortstack{Emotion}} & \multirow{2}{*}{\shortstack{CBA}} & ASR & 1.000 & 0.211 & 0.132 & 0.225 & 0.218 & \textbf{0.045} & 0.105 &0.063 & 1.000 & 0.893 & 0.166 & 0.318 & 0.153 & 0.969 &0.092& \textbf{0.082} \\
        && CDA & 0.938 & 0.369 & 0.433 & 0.623 & 0.813 & 0.511 & 0.725 &\textbf{0.902} & 0.936 & 0.879 & 0.278 & 0.881 &0.916  &0.903 & 0.597&\textbf{0.918} \\

        \cmidrule{2-19}
        & \multirow{2}{*}{\shortstack{BadEdit}} & ASR & 0.821 & 0.257 & 0.132 & 0.106 & 0.212 & 0.779 &0.182 &\textbf{0.101} & 1.000 & 0.163 & \textbf{0.001} & 0.732 & 0.274 & 0.727 &0.208& 0.153 \\
        && CDA & 0.534 & 0.300 & 0.420 & 0.516 & 0.491 & 0.471 &0.375 &\textbf{0.512} & 0.607 & 0.353 & 0.085 & 0.339 & 0.538 & 0.376 & 0.383&\textbf{0.577}\\
        \cmidrule{2-19}
        

        & \multirow{2}{*}{\shortstack{VPI}} & ASR & 0.988 & 0.597 & 0.140 & 0.175 & 0.128 & 0.352 & 0.184&\textbf{0.117} & 1.000 & 0.687 & \textbf{0.069} & 0.757 & 0.256 & 0.115 &0.171& 0.181  \\
        && CDA & 0.949 & 0.407 & 0.466 & 0.828 & \textbf{0.883} & 0.554 &0.632 &{\color{black}0.861} & 0.950 & 0.856 & 0.351 & 0.813 & 0.731 & 0.600 & 0.665&\textbf{0.926} \\

        \cmidrule{2-19}
        & \multirow{2}{*}{\shortstack{BadNet}} & ASR & 1.000 & 0.741 & 0.184 & 0.550 & 0.114 & 0.355 & 0.294&\textbf{0.108} & 0.333 & 0.145 & 0.141 & 0.595 & 0.210 & 0.356 &0.057& \textbf{0.128}\\
        && CDA & 0.887 & 0.830 & 0.489 & 0.805 & 0.808 & 0.914 & 0.683&\textbf{0.868} & 0.882 & 0.890 & 0.393 & 0.837 & \textbf{0.891} & 0.607 & 0.674&0.868 \\
             \midrule
         \multirow{10}{*}{\shortstack{SST-2}} & 
        
        \multirow{2}{*}{\shortstack{CBA}} & ASR & 1.000 & 0.744 & 0.099 & 0.496 & 0.216 & 0.292 & 0.578 &\textbf{0.044} & 0.682 & 0.136 & 0.099 & 0.489 & 0.197 & 0.361 & 0.759&\textbf{0.165}\\
        && CDA & 0.957 & 0.872 & 0.581 & 0.928 & 0.826 & 0.621 & \textbf{0.935}&0.913& 0.925 & 0.923 & 0.581 & 0.827 & \textbf{0.917} & 0.848 &0.879& 0.893\\

        \cmidrule{2-19}

        & \multirow{2}{*}{\shortstack{BadEdit}} & ASR & 0.932 & 0.630 & 0.118 & 0.216 & 0.100 & 0.157 & 0.191&\textbf{0.093} & 0.974 & 0.161 & 0.118 & 0.813 & 0.161 & 0.575 &0.131 &\textbf{0.096} \\
        && CDA & 0.951 & 0.724 & 0.733 & 0.823 & 0.873 & 0.692 & 0.798&\textbf{0.925} & 0.951 & 0.658 & 0.733 & 0.848 & \textbf{0.880} & 0.799 & 0.405&\color{black}0.855\\

        \cmidrule{2-19}

        & \multirow{2}{*}{\shortstack{VPI}} & ASR & 1.000 & 0.387 & 0.240 & 0.495 & 0.151 & \textbf{0.112} &0.176& 0.163 & 1.000 & 0.489 & 0.240 & 0.850 & 0.236 & 0.274 & 0.330&\textbf{0.220}\\
        && CDA & 0.964 & 0.555 & 0.765 & 0.925 & 0.693 & 0.856 &0.926 &\textbf{0.938} & 0.968 & 0.923 & 0.765 & 0.854 & 0.895 & 0.756 &0.928 &\textbf{0.929} \\

        \cmidrule{2-19}

        & \multirow{2}{*}{\shortstack{BadNet}} & ASR & 1.000 & 0.848 & 0.247 & 0.503 & 0.108 & 0.490 & 0.191&\textbf{0.093} & 1.000 & 0.147 & 0.247 & 0.369 & 0.146 & 0.501 & 0.152&\textbf{0.107} \\
        && CDA & 0.954 & 0.831 & 0.774 & 0.955 & 0.852 & \textbf{0.940} &0.935& 0.925 & 0.914 & 0.876 & 0.774 & 0.931 & 0.820 & \textbf{0.928} & 0.913&0.870 \\

        \midrule
         \multirow{10}{*}{\shortstack{COLA}} & 
        \multirow{2}{*}{\shortstack{CBA}} & ASR & 1.000 & 0.799 & 0.483 & 0.433 & \textbf{0.081} & 0.571 &0.366 &0.097 & 1.000 & 0.939 & 0.288 & 0.376 & 0.174 & 0.216 &0.399 &\textbf{0.154}\\
        && CDA & 0.901 & 0.695 & 0.795 & 0.836 & 0.657 & 0.745 & \textbf{0.910}&0.875 & 0.881 & 0.855 & \textbf{0.879} & 0.862 & 0.877 & 0.877 & 0.449&0.863\\

        \cmidrule{2-19}        

        & \multirow{2}{*}{\shortstack{BadEdit}} & ASR & 1.000 & 0.157 & 0.470 & 0.591 & 0.101 & 0.678 & 0.209&\textbf{0.081} & 0.901 & 0.102 & 0.294 & 0.081 & 0.214 & \textbf{0.000} & 0.240 &0.089 \\
        && CDA & 0.721 & 0.319 & 0.792 & 0.806 & 0.671 & 0.691 & 0.731&\textbf{0.742} & 0.748 & 0.309 & \textbf{0.817} & 0.610 & 0.675 & 0.309 &0.654& 0.719 \\

        \cmidrule{2-19}

        & \multirow{2}{*}{\shortstack{VPI}} & ASR & 1.000 & 0.239 & 0.409 & 0.486 & 0.154 & \textbf{0.000} &0.313& 0.150 & 0.931 & 0.552 & 0.849 & 0.852 & 0.164 & 0.195 & 0.470&\textbf{0.157} \\
        && CDA & 0.916 & 0.815 & 0.764 & 0.837 & 0.791 & 0.763 & 0.813&\textbf{0.825} & 0.932 & 0.654 & 0.833 & 0.847 & 0.873 & 0.742 &0.797& \textbf{0.878} \\

        \cmidrule{2-19}
        
        & \multirow{2}{*}{\shortstack{BadNet}} & ASR & 1.000 & 0.796 & 0.371 & 0.486 & 0.197 & 0.121 & 0.003&\textbf{0.027} & 1.000 & 0.534 & 0.870 & 0.691 & 0.197 & 0.213 & 0.661&\textbf{0.175} \\
        && CDA & 0.823 & 0.751 & 0.715 & 0.810 & 0.750 & 0.719 & 0.703&\textbf{0.813} & 0.801 & 0.825 & 0.847 & 0.745 & 0.767 & 0.719 & 0.750&\textbf{0.771}\\
        \midrule
        \multirow{10}{*}{\shortstack{MNLI}} 
        & \multirow{2}{*}{\shortstack{CBA}} & ASR & 0.997 & 0.645 & 0.225 & 0.421 & 0.117 & 0.154 & \textbf{0.054}& 0.140& 0.997 & 0.882 & 0.154 & 0.344 & 0.171 & 0.776 &0.012& \textbf{0.149} \\
        && CDA & 0.898 & 0.442 & 0.146 & 0.746 & 0.872 & 0.410 & 0.824 &\textbf{0.869} & 0.899 & \textbf{0.879} & 0.117 & 0.889 & 0.762 & 0.866 &0.252& 0.868 \\

        \cmidrule{2-19}

       & \multirow{2}{*}{\shortstack{BadEdit}} & ASR & 0.735 & 0.121 & 0.133 & 0.075 & 0.169 & 0.720 &0.061& \textbf{0.101} & 0.968 & 0.188 & \textbf{0.007} & 0.572 & 0.029 & 0.881&0.093 & 0.041\\
        && CDA & 0.446 & 0.447 & 0.323 & 0.727 & 0.462 & 0.447 & 0.235 &\textbf{0.643} & 0.530 & 0.452 & 0.243 & 0.459 & 0.387 & 0.454 &0.565& \textbf{0.497} \\
        
        \cmidrule{2-19}

        & \multirow{2}{*}{\shortstack{VPI}} & ASR & 1.000 & 0.043 & 0.572 & 0.221 & 0.065 & \textbf{0.000} & 0.506& 0.085 & 1.000 & 0.826 & 0.378 & 0.364 & \textbf{0.272} & 0.435 &0.501& 0.331 \\
        && CDA & 0.470 & 0.306 & 0.298 & 0.578 & 0.457 & \textbf{0.584} &0.441& 0.458 & 0.380 & 0.362 & 0.196 & 0.869 & 0.375 & \textbf{0.578} & 0.412&0.462 \\

        \cmidrule{2-19}
        & \multirow{2}{*}{\shortstack{BadNet}} & ASR & 1.000 & 0.179 & 0.194 & 0.345 & 0.083 & 0.361 & 0.314&\textbf{0.016} & 1.000 & 0.831 & 0.062 & 0.546 & 0.100 & 0.063 &0.647& \textbf{0.056}\\
        && CDA & 0.866 & 0.055 & 0.427 & 0.873 & 0.779 & \textbf{0.863} & 0.827& 0.831 & 0.827 & 0.675 & 0.036 & 0.822 & 0.678 & 0.313 & 0.636&\textbf{0.817}\\
        \midrule

        \multirow{10}{*}{\shortstack{QQP}} & \multirow{2}{*}{\shortstack{CBA}} & ASR & 1.000 & 0.698 & 0.141 & 0.388 & 0.174 & 0.116 & \textbf{0.057}& 0.139 & 1.000 & 0.964 & \textbf{0.026} & 0.573 & 0.170 & 0.674 &0.801& 0.148\\
        && CDA & 0.935 & \textbf{0.971} & 0.020 & 0.434 & 0.891 & 0.689 & 0.850 & 0.905 & 0.929 & 0.514 & 0.010 & 0.858 & 0.775 & 0.878 &0.421& \textbf{0.897} \\

        \cmidrule{2-19}

        & \multirow{2}{*}{\shortstack{BadEdit}} & ASR & 0.969 & \textbf{0.042} & 0.312 & 0.200 & 0.153 & 0.804 &0.066& 0.082 & 0.978 & 0.076 & \textbf{0.027} & 0.627 & 0.199 & 0.799 & 0.079&0.089\\
        && CDA & 0.600 & 0.619 & 0.536 & 0.632 & 0.518 & 0.616 & 0.095&\textbf{0.694} & 0.702 & 0.516 & 0.057 & 0.611 & 0.484 & 0.602 & 0.273&\textbf{0.671}\\

        \cmidrule{2-19}

        & \multirow{2}{*}{\shortstack{VPI}} & ASR & 1.000 & 0.638 & \textbf{0.075 }& 0.190 & 0.095 & 0.315 &0.435& 0.136 & 1.000 & 0.309 & \textbf{0.053} & 0.859 & 0.239 & 0.633 &0.595& \textbf{0.200} \\
        && CDA & 0.912 & 0.486 & 0.216 & 0.625 & 0.783 & 0.553 & 0.812&\textbf{0.870} & 0.922 & 0.468 & 0.146 & 0.802 & \textbf{0.900} & 0.588 & 0.794&0.889 \\

        \cmidrule{2-19}
        & \multirow{2}{*}{\shortstack{BadNet}} & ASR & 1.000 & 0.752 & 0.185 & 0.530 & \textbf{0.137} & 0.671 & 0.858 &0.177 & 1.000 & 0.726 & \textbf{0.065} & 0.673 & 0.094 & 0.235 &0.829&0.132 \\
        && CDA & 0.853 & 0.700 & 0.382 & 0.818 & \textbf{0.871} & 0.389 &0.376& 0.812 & 0.847 & 0.825 & 0.192 & 0.841 & 0.830 & 0.389 &0.800&\textbf{0.869}\\

        \bottomrule
    \end{tabular}

\end{table*}

\smallskip
\noindent\textbf{Baselines.} We compare \sys against 6 representative backdoor defense methods. Wanda (WAN) \cite{sun2023simple} is a lightweight, training-free baseline that prunes parameters with low activation-aware importance scores. Fine-Tuning (F/T) \cite{qi2023fine} performs retraining on clean data to overwrite backdoor behaviors. While widely adopted, this strategy often leads to degraded model utility.  Neural Attention Distillation (NAD) \cite{li2021neural} applies a distillation-based fine-tuning scheme, where a clean teacher model regularizes intermediate representations of the backdoored model. BEEAR (BEE) \cite{zeng2024beear} mitigates backdoors by leveraging the uniform drifts induced by backdoor behaviors in the model's embedding space. CROW \cite{min2024crow} enforces layer-wise representation consistency through fine-tuning with internal regularization. 
LETHE (LET) \cite{chen2025lethe} introduces knowledge dilution by training a lightweight clean model for model merging, together with benign, semantically relevant evidence prompting.
These baselines reflect state-of-the-art purification methods in the literature.

Additional experimental details are shown in the Appendix.

\subsection{Main Results}

We conduct extensive comparisons between \sys and 6 representative defenses (WAN, F/T, NAD, BEE, CROW, and LET) across two widely-used LLMs (Llama3-8B and Mistral-7B). The results for \emph{classification} and \emph{generation} tasks are summarized in Table~\ref{tab:main-result} and Table~\ref{tab:main-result2}, respectively. To avoid clean-data leakage, we strictly exclude any prototypes constructed using the evaluation dataset during purification. 

\noindent\textbf{Classification Tasks.}
Across 5 classification datasets and 4 representative attacks, \sys consistently achieves a strong purification-utility trade-off. Overall, it delivers the most stable performance among the evaluated baselines.

\sys remains effective against single-trigger attacks, including the conventional poisoning (BadNet) and the more challenging model-editing-based BadEdit. On SST-2 under BadEdit, \sys attains ASR 9.3\%/9.6\% with CDA 92.5\%/85.5\% (Llama3-8B/Mistral-7B), outperforming WAN (ASR 63.0\%/16.1\%) and NAD (ASR 21.6\%/81.3\%), while avoiding the large utility degradation observed with F/T (CDA 73.3\%/73.3\%). Under BadNet on MNLI, \sys further reduces ASR to 1.6\%/5.6\% while retaining CDA 83.1\%/81.7\%, exceeding all baselines by a clear margin.

For multi-trigger attack CBA, \sys consistently reduces ASR to below 16.5\% with <3\% CDA degradation, whereas other defenses typically exhibit either higher residual ASR or larger utility loss. On SST-2 with Llama3-8B, WAN and NAD reduce ASR from 100\% only to 74.4\% and 49.6\%, respectively. While F/T, BEE, and CROW can suppress ASR to 9.9\%, 21.6\%, and 29.2\%, they do so at the cost of clean utility (CDA 58.1\% for F/T and 62.1\% for CROW). In contrast, \sys achieves ASR 4.4\% while maintaining a substantially higher CDA 91.3\%.


\sys also remains effective against triggerless attack VPI. On Emotion (Llama3-8B), it reduces ASR from 98.8\% to 11.7\% while maintaining CDA 86.1\%, outperforming WAN (ASR 59.7\%, CDA 40.7\%), F/T (ASR 14.0\%, CDA 46.6\%), NAD (ASR 17.5\%, CDA 82.8\%), BEE (ASR 12.8\%, CDA 88.3\%), and CROW (ASR 35.2\%, CDA 55.4\%).



\begin{table}
    \renewcommand{\arraystretch}{0.7}
    \caption{
    Performance of \sys on Generation Tasks.    
    }
    \label{tab:main-result2}
    \centering
    \scriptsize
    \setlength\tabcolsep{1pt}
    \begin{tabular}{lc|cccccccc}
        \toprule
        \multicolumn{10}{c}{\textbf{Llama3-8B}} \\
        \midrule
        Attack & Metrics & Backdoor & WAN & F/T & NAD & BEE & CROW  & LET & \textbf{\sys} \\
        \midrule
        \multirow{2}{*}{\shortstack{DTBA}} & ASR & 0.595 & 0.695 & 0.100 & 0.490 & 0.105 & \textbf{0.035} & 0.255 & 0.085 \\
                                        & CDA & 0.810 & 0.330 & \textbf{0.880} & 0.820 & 0.790 & 0.490 & 0.820 & 0.805 \\
        \cmidrule{1-10}
        \multirow{2}{*}{\shortstack{AutoPoison}} & ASR & 0.800 & 0.067 & 0.210 & 0.618 & 0.050 & 0.227 & 0.143 & \textbf{0.065}  \\
                                             & CDA & 0.999 & 0.572 & 0.908 & 0.905 &0.950  & 0.806 & 0.921 & \textbf{0.955}  \\
        \cmidrule{1-10}
        \multirow{2}{*}{\shortstack{VPI}} & ASR & 0.990 & 0.230 & 0.190 & 0.105 &0.095  & 0.150 & 0.120 & \textbf{0.080}  \\
                                        & CDA & 0.975 & 0.275 & 0.795 & 0.890 &0.925  & 0.820 & 0.870 & \textbf{0.945} \\
        \midrule
        \multicolumn{10}{c}{\textbf{Mistral-7B}} \\
        \midrule
        Attack & Metrics & Backdoor & WAN & F/T & NAD & BEE & CROW & LET &\sys \\
        \midrule
        \multirow{2}{*}{\shortstack{DTBA}} & ASR & 0.770 & 0.735 & 0.130 & 0.765 & \textbf{0.050}  & 0.525 & 0.685& 0.060  \\
                                        & CDA & 0.550 & 0.290 & \textbf{0.960} & 0.590 & 0.830 & 0.720 & 0.590& 0.875 \\
        \cmidrule{1-10}
        \multirow{2}{*}{\shortstack{AutoPoison}} & ASR & 0.827 & 0.775 & 0.083 & 0.814 &0.110  & 0.266 & 0.062 & \textbf{0.080} \\
                                             & CDA & 0.999 & 0.901 & 0.824 & 0.906 & \textbf{0.960} & 0.835 & 0.890 & \textbf{0.960} \\
        \cmidrule{1-10}
        \multirow{2}{*}{\shortstack{VPI}} & ASR & 1.000 & 0.560 & 0.092 & 0.902 &0.280  & 0.255 & 0.325 & \textbf{0.090}  \\
                                        & CDA & 0.990 & 0.735 & 0.750 & 0.935 & 0.920 & 0.785 & 0.885 & \textbf{0.950}\\
        \bottomrule
    \end{tabular}
\end{table}

\noindent\textbf{Generation Tasks.}
We further evaluate \sys on the \textsc{Chat-backdoor} generation benchmark against 3 representative attacks, i.e., DTBA, AutoPoison, and VPI. Notably, this evaluation is based on the prototypes trained on the classification tasks. 
As shown in Table~\ref{tab:main-result2}, \sys consistently achieves low ASR and high CDA on both Llama3-8B and Mistral-7B, demonstrating the transferability of the prototypes across the task domains.
For \textsc{DTBA}, \sys reduces ASR from 59.5\%/77.0\% to 8.5\%/6.0\% (Llama3-8B/Mistral-7B), while maintaining CDA at 80.5\%/87.5\%.
Under AutoPoison, \sys attains ASR of 6.5\%/8.0\% with CDA of 95.5\%/96.0\%.
For VPI, \sys manages to suppress ASR to 8.0\%/9.0\% and retain CDA of 94.5\%/95.0\%. 
In contrast, prior defenses are either ineffective (e.g., WAN, NAD, and LET) or incur notable utility degradation (e.g., WAN and CROW). 
Notably, WAN can even increase ASR after defense, likely because editing with non-safety-aligned data disrupts safety-critical representations and induces additional unsafe outputs.
While F/T is competitive on DTBA, it exhibits a more pronounced utility drop under VPI. For instance, CDA decreases from 97.5\% to 79.5\% on Llama3-8B and from 99.0\% to 75.0\% on Mistral-7B.
Among baselines, BEE is relatively robust but remains inferior to \sys in both ASR reduction and CDA preservation.


\subsection{Ablation Study}
This section presents an ablation study to evaluate the contribution of each component in \sys. The experiments are conducted using Llama-3 as the base model.
\begin{table}[t]
\centering
\caption{Comparison of Backdoor Vector Construction Strategies on Purification Performance.}
\label{tab:ablation-study-for-benign-vector}
\renewcommand{\arraystretch}{1.1}
\setlength\tabcolsep{2pt}
\scriptsize
\begin{tabular}{cc|cccc|cccc}
\toprule
\multirow{2}{*}{\textbf{Method}} & \multirow{2}{*}{\textbf{Metrics}} 
& \multicolumn{4}{c|}{\textbf{Emotion}} 
& \multicolumn{4}{c}{\textbf{SST-2}} \\
\cmidrule(lr){3-6} \cmidrule(lr){7-10}
 &  & CBA & BadEdit & VPI & BadNet  & CBA & BadEdit & VPI & BadNet \\
\midrule
\multirow{2}{*}{$v_b$} & ASR &0  & 0.076 &0  & 0.231 &0.521  &  0.459& 0.599 & 0.735 \\
                   & CDA & 0.060 & 0.244 & 0 & 0.155 & 0.834 & 0.818 &0.806  & 0.790 \\
\multirow{2}{*}{\makecell{$v_b - v_c$\\(ours)}} & ASR &0.063  & 0.101  & 0.117 & 0.108 & 0.044 &0.093  &0.163  & 0.093  \\
                    & CDA & 0.902  &0.492  &0.461  &0.868  &0.891  &0.859  &0.880  &0.905  \\

\bottomrule
\end{tabular}
\end{table}

\smallskip
\noindent\textbf{Impact of Benign Vector Removal.} 
We first examine the design choice of defining the backdoor vector as the difference between malicious and benign task vectors (e.g., $v_b - v_c$). As discussed in Section \ref{sec:backdoor-simulation}, this subtraction is intended to remove task-specific signals. We evaluate the effect of this removal by comparing our default construction $v_b - v_c$ against the variant that directly uses $v_b$, and present the results in Table \ref{tab:ablation-study-for-benign-vector}.  

The comparison clearly demonstrates the importance of removing the benign task vector. When directly using $v_b$, the ASR cannot be effectively reduced, while CDA deteriorates substantially across most settings. This effect is particularly severe on the Emotion dataset, where $v_b$-based purification leads to model failure, with both ASR and CDA collapsing to zero. These results support our hypothesis that $v_b$ inherently entangles malicious backdoor-induced updates with benign task-related updates. Consequently, the prototype derived from $v_b$ suppresses both malicious and benign signals in subsequent purification, leading to substantial utility loss.


\smallskip
\noindent\textbf{Impact of Aggregation Methods.} 
We compare the effects of AM and PCA aggregation strategies on backdoor mitigation performance, and summarize their results in Table \ref{tab:ablation-study-aggregation-methods}. Overall, the simpler AM-based aggregation consistently outperforms PCA. On average, AM-based \sys reduces the ASR to 9.7\% and 9.8\%, while maintaining CDA at 78.1\% and 88.4\% on the Emotion and SST-2 datasets, respectively. In contrast, PCA-based aggregation yields notably higher ASR values (17.0\% and 17.2\%) and incurs larger CDA degradation, with accuracies dropping to 71.4\% and 77.8\%. This performance gap is likely due to the misaligned objective of PCA. Specifically, PCA identifies directions that maximize variance across the backdoor vectors, which do not necessarily align with the shared backdoor-related signal. As a result, PCA may amplify dataset-specific or attack-irrelevant variations, leading to a noisy prototype. 
Moreover, PCA operates on mean-centered vectors, removing consistent mean-shift signals that are critical for subsequent boundary layer detection and model purification. 
Therefore, we adopt AM as the default aggregation function in the \sys framework.


\smallskip
\noindent\textbf{Impact of Boundary Layer Selection.} 
\begin{figure}[t]
    \centering
    \includegraphics[width=\columnwidth]{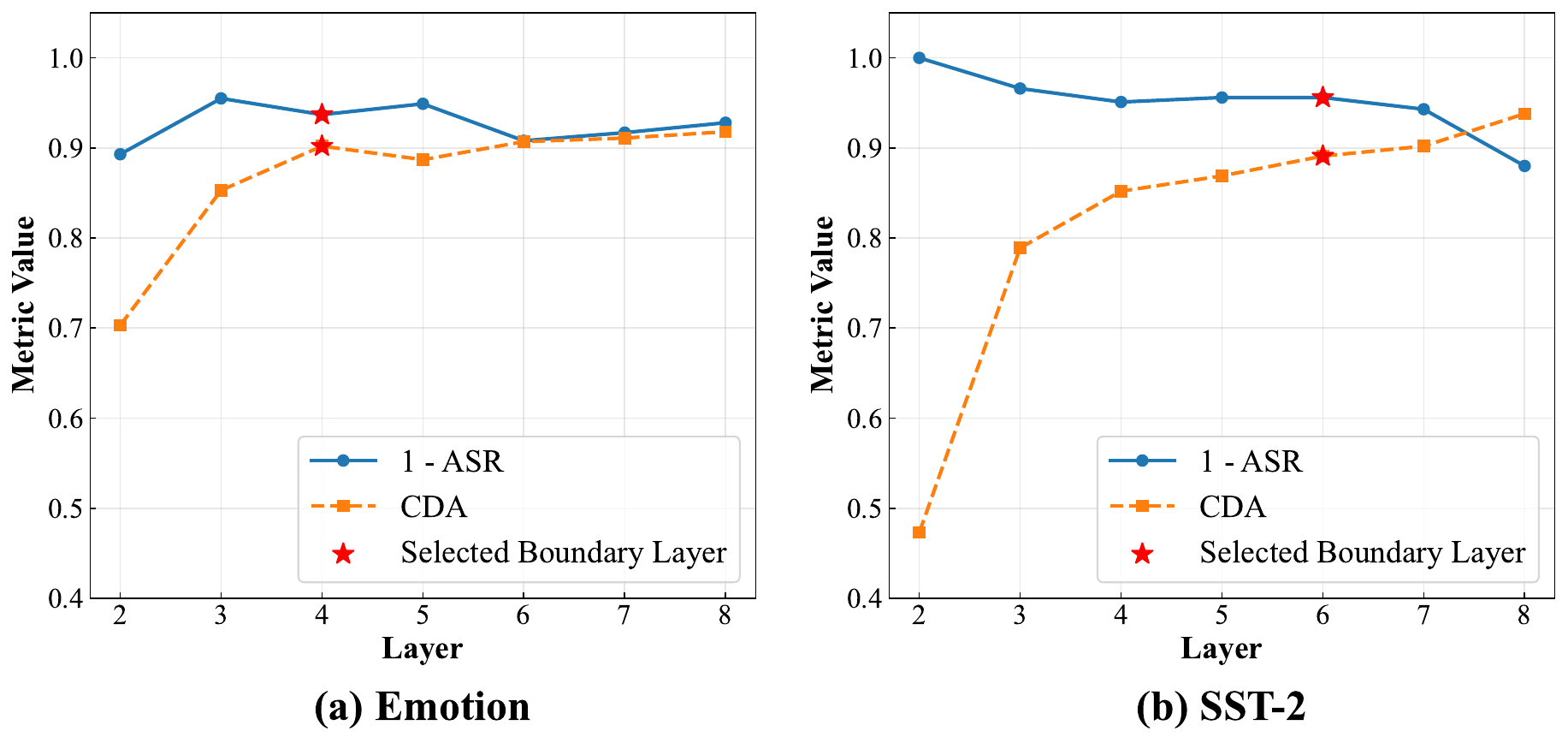}
    \caption{Effect of Boundary Layer Detection.} 
    \label{fig:boundary-layer-detection}
\end{figure}
We analyze the effect of the boundary layer selection module by evaluating the performance of \sys under different choices of boundary layers. The results are reported in Table \ref{fig:boundary-layer-detection}. As the boundary layer moves deeper, a larger portion of the lower layers is protected, leading to an increase in CDA. Meanwhile, (1 - ASR) decreases, since fewer layers participate in purification. Our significance-based boundary layer selection mechanism identifies layer 4 for the Emotion dataset and layer 6 for SST-2, achieving ASR values of 6.3\% and 4.4\%, and CDA values of 90.2\% and 89.1\%, respectively. The selected layers exhibit an effective trade-off between attack success reduction and clean accuracy preservation.

\begin{figure}[tt]
    \centering
    \includegraphics[width=\columnwidth]{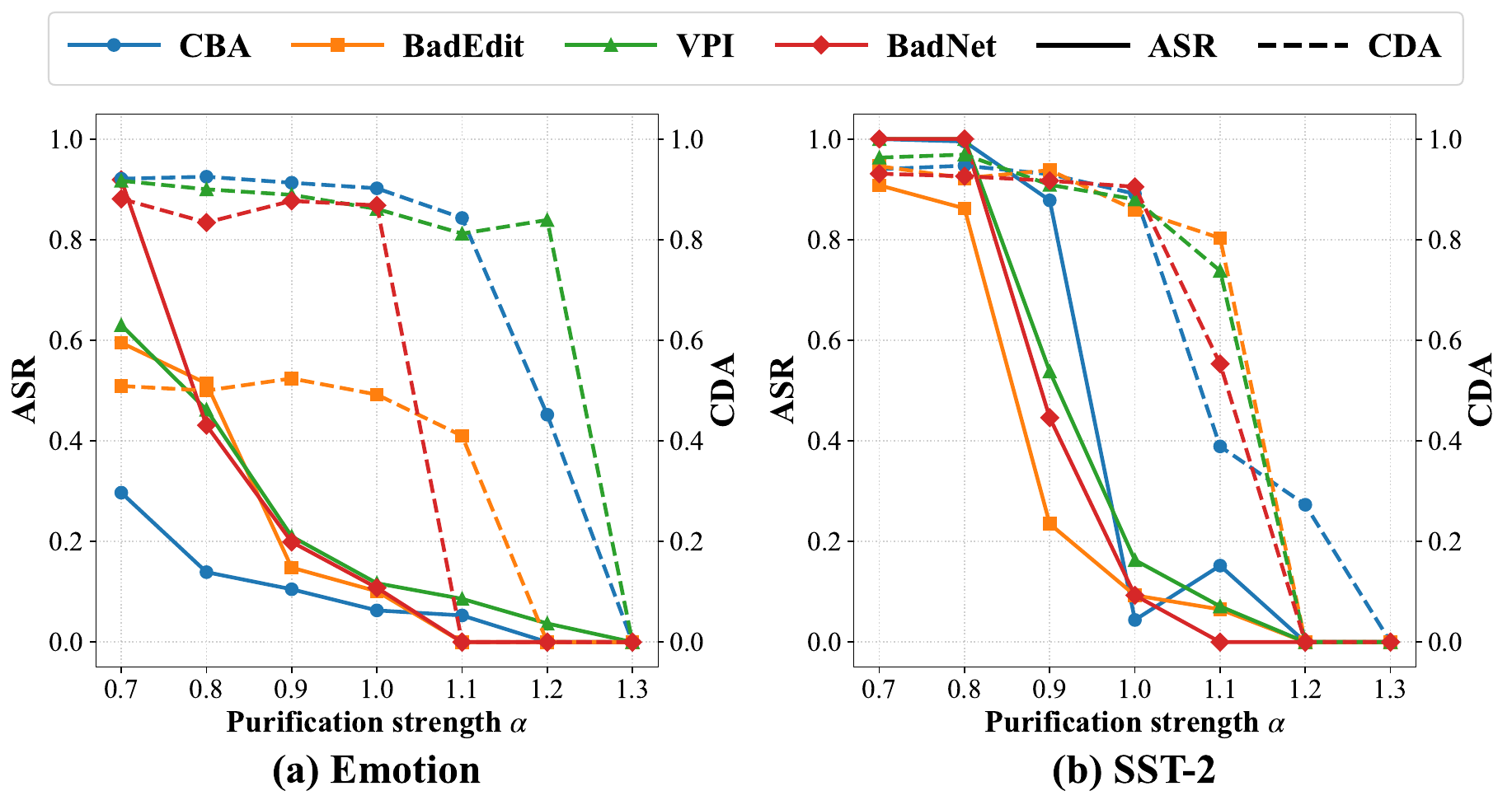}
    \caption{Effect of purification strength $\alpha$ .} 
    \label{fig:purification-strength}
\end{figure}
\smallskip
\noindent\textbf{Impact of Purification Strength.} 
We further investigate how the purification strength $\alpha$ affects the trade-off between backdoor mitigation and model utility. 
While $\alpha \in [0, 1]$ naturally controls the extent of backdoor suppression, we additionally explore values beyond this range to study the consequences of over-purification. The results are summarized in Table \ref{fig:purification-strength}. Unsurprisingly, ASR consistently decreases across both datasets as $\alpha$ increases. However, this improvement comes at the cost of clean performance. When $\alpha$ approaches 1.0, ASR drops drastically, while CDA remains relatively high. When $\alpha > 1.0$, CDA rapidly collapses to zero. Overall, setting $\alpha$ within the range of $[0.9, 1.0]$ achieves a reasonable balance. This observation provides defenders a practical heuristic for selecting purification strength when deploying \sys in real-world scenarios.

\section{Discussion}

\subsection{Backdoor Purification as a Service} 
\sys naturally supports BDaaS, where potentially compromised LLMs can be submitted for inspection and purification. In the following, we discuss how \sys fulfills the requirements of BDaaS, i.e.,  reusability, customizability, interpretability, and runtime efficiency.

\smallskip
\noindent\textbf{Reusability.} 
Our experimental results in Table \ref{tab:main-result} and Table \ref{tab:main-result2} demonstrate that \sys exhibits strong reusability. In particular, the backdoor vector pool is constructed once through offline simulation using auxiliary datasets and known attack strategies. This pool can then be reused to effectively purify multiple target backdoored models that share the same architecture, even when they are trained on different datasets, task domains, and under a variety of backdoor attacks.

\smallskip
\noindent\textbf{Customizability.} 
A practical BDaaS platform should support customized usage scenarios. Specifically, when partial knowledge about the backdoor, e.g., the anticipated attack family or the potential downstream task, is available, the platform should be able to leverage this knowledge to improve purification effectiveness without introducing significant computational overhead. To evaluate this capability, we consider two representative scenarios, i.e., attack-aware and task-aware settings. In this study, the defender is given coarse-grained knowledge of candidates, rather than the exact attack or task.

In the attack-aware setting, the defender is provided with the attack category $\mathcal{C}_{A} = \{A^{(i)}\}_{i=1}^{n_a}$, which is likely used to compromise the target model. Accordingly, we conduct prototype construction using 10 simulated backdoor vectors corresponding to the trigger types of attacks $A^{(i)} \in \mathcal{C}_{A}$. As a baseline, we consider a random-selection strategy that samples 10 backdoor vectors, regardless of attack type. The results of targeted purification on CBA and BadEdit are reported in Table \ref{tab:discussion-customizability-trigger} (Appendix). Compared to random selection, the attack-aware strategy consistently achieves lower ASR while preserving higher CDA across both attacks and datasets. 

In the task-aware setting, we assume that the defender has prior knowledge of the potential downstream task set $\mathcal{C}_{D} = \{D^{(i)}\}_{i=1}^{n_d}$. We first construct a task-related shadow dataset $\widehat{D}$ based on this information, and then evaluate \sys on $\widehat{D}$ by comparing purification performance under two backdoor vector pool conditions: (i) a pool that includes backdoor vectors derived from $\mathcal{C}_{D}$, and (ii) a pool constructed without any task-related backdoor vectors. The corresponding results are summarized in Table \ref{tab:discussion-customizability-task} (Appendix). Incorporating task-related vectors consistently improves purification effectiveness, producing ASR reductions of 8.0\% and 3.4\%, along with CDA improvements of 4.0\% and 3.3\% on Emotion and SST-2, respectively. These results demonstrate that auxiliary knowledge about the anticipated attacks or downstream tasks can be effectively exploited by \sys to further improve backdoor mitigation.

\begin{table}[t]
\centering
\caption{Analysis of Backdoor Signal for the Clean and Compromised Models.}
\label{tab:backdoor-signal}
\renewcommand{\arraystretch}{1.0}
\setlength\tabcolsep{4pt}
\scriptsize
\begin{tabular}{clccc}
\toprule
No. & Granularity & Clean & CBA ($\Delta_{\text{CBA}}$)& BadEdit($\Delta_{\text{BadEdit}}$) \\

\specialrule{0.08em}{0.4ex}{0.4ex}
\specialrule{0.08em}{0ex}{0.6ex}
\multicolumn{2}{l}{\textbf{\emph{\footnotesize{All}}} }\\
1 & All & $0.491_{\pm 0.147}$ & $0.540_{\pm 0.102} (+0.049)$ & $2.108_{\pm 0.128} (1.617)$\\
\specialrule{0.08em}{0.4ex}{0.4ex}
\specialrule{0.08em}{0ex}{0.6ex}
\multicolumn{2}{l}{\textbf{\emph{\footnotesize{Blocks}}} }\\
2 & Self-Attention & $0.561_{\pm 0.187} $ & $0.565_{\pm 0.149} (+0.004)$ & $2.151_{\pm 0.151} (+1.590)$\\
3 & MLP & $0.398_{\pm 0.006}$ & $0.505_{\pm 0.030} (+0.107)$ & $2.051_{\pm 0.133} (+1.653)$\\
\specialrule{0.08em}{0.4ex}{0.4ex}
\specialrule{0.08em}{0ex}{0.6ex}
\multicolumn{2}{l}{\textbf{\emph{\footnotesize{Projections}}} }\\

4 & $Q_{\text{proj}}$ & $0.746_{\pm 0.017}$ & $ 0.727_{\pm 0.019} (-0.019)$ & $2.204_{\pm 0.166} (+1.458)$\\
5 & $K_{\text{proj}}$ & $0.372_{\pm 0.014}$ & $0.417_{\pm 0.027} (+0.045)$ & $2.260_{\pm 0.125} (+1.888)$\\
6 & $V_{\text{proj}}$ & $0.374_{\pm 0.009}$ & $ 0.414_{\pm 0.018} (+0.040)$ & $2.105_{\pm 0.112} (+1.731)$\\
7 & $O_{\text{proj}}$ & $0.751_{\pm 0.012}$ & $ 0.703_{\pm 0.039} (-0.048)$ & $2.034_{\pm 0.135} (+1.283)$\\
8 & $\text{Up}_{\text{proj}}$ & $0.402_{\pm 0.007}$ & $0.497_{\pm 0.027} (+0.095)$ & $2.133_{\pm 0.126} (+1.731)$\\
9 & $\text{Gate}_{\text{proj}}$ & $0.397_{\pm 0.005}$ & $0.508_{\pm 0.023} (+0.111)$ & $2.077_{\pm 0.091} (+1.680)$\\
10 & $\text{Down}_{\text{proj}}$ & $0.396_{\pm 0.006}$ & $0.511_{\pm 0.041} (+0.115)$ & $1.944_{\pm 0.136} (+1.548)$\\

\bottomrule
\end{tabular}
\end{table}
\smallskip
\noindent\textbf{Interpretability.}
We analyze the backdoor signals $c_i$ identified by \sys, which provide an interpretable insight into where and how backdoor updates are encoded in a submitted model. Table \ref{tab:backdoor-signal} compares clean fine-tuned models with models compromised by a data-poisoning attack (CBA) and a weight-poisoning attack (BadEdit). Overall, both attacks exhibit enhanced backdoor signals relative to the clean baseline, but with different magnitudes. CBA produces only a modest increase ($\Delta_{\text{CBA}} = +0.049$), whereas BadEdit induces a much stronger effect ($\Delta_{\text{BadEdit}} = +1.617$), roughly two order of magnitude larger. Empirically, these signal statistics may help diagnose whether a submitted model is compromised and, if so, provide a broad hint about the likely attack family.

To further localize the signal, we decompose $c_i$ by blocks and projections. At the sub-layer level, the increase concentrates primarily in MLP blocks ($+0.107$) rather than self-attention ($+0.004$), indicating that backdoor updates are predominantly stored in MLP pathways (about $27\times$ in our measurement). This observation is aligned with the findings from prior work \cite{li2024badedit, meng2022locating}. At the projection level within attention, we observe small positive shifts in $K_\text{proj}$ and $V_\text{proj}$, while $Q_\text{proj}$ and $O_\text{proj}$ exhibit weaker or even negative changes. This observation indicates that the backdoor is preferentially injected into projections that store the ``content'' (i.e., Keys and Values), whereas Query and Output projections may be shaped more by benign task adaptation. Meanwhile, all MLP projections (Up/Gate/Down) exhibit consistent increases. These projection-wise patterns motivate a practical extension: defenders could apply adaptive purification strength across blocks (e.g., stronger suppression on MLP than attention, or module-specific calibration within attention) to better balance mitigation and utility. We leave this direction for future work.


\smallskip
\noindent\textbf{Runtime Efficiency.}
We decompose \sys into a one-time offline phase and a per-model service-time phase. Specifically, prototypes are constructed offline (Stage I) and then reused to process submitted models at service time via three lightweight stages, i.e., prototype construction (Stage II), boundary layer detection (Stage III), and model purification (Stage IV). Because the service-time pipeline requires no per-model training, \sys process each model in 5-10 minutes. 
In contrast, training-based baselines incur substantial per-model overhead. For example, defenses such as F/T, NAD, and LET rely on iterative gradient updates for every target model, typically requiring
hundreds of minutes for training \cite{chen2025lethe}. Overall, once prototypes are prepared offline, \sys supports fast, high-throughput purification in the runtime.

\subsection{Adaptive Attack}
We consider an adaptive adversary who is aware of \sys and attempts to preserve backdoor effectiveness after purification.
Specifically, the adversary first estimates a backdoor prototype and then identifies the corresponding backdoored components. Using this information, the adversary pre-amplifies the backdoor signal $\lambda_i$ {\color{black}(Eq. (\ref{eq:backdoor-mitigation}))} to reduce the effect of suppression introduced by prototype-based purification. We analyze two attacker knowledge settings: (i) \emph{prototype not leaked}, where the attacker knows only the defense strategy but not the exact prototype $p^{*}$ used for purification; and (ii) \emph{prototype leaked}, where the defender's prototype $p^{*}$ is exposed to the attacker.

We initially experimented with stronger amplification ($1.5\lambda_i$ and $2\lambda_i$) on Emotion under the CBA attack. However, both settings severely degraded utility: $1.5\lambda_i$ reduces CDA to 48.3\%, while $2\lambda_i$ leads to incoherent generation. We therefore adopt a milder amplification factor of $1.2\lambda_i$, which provides the best balance between the backdoor persistence and utility. Table \ref{tab:discussion-adaptive-attack} (Appendix) summarizes the results on Emotion and SST-2 datasets. Under ``Prototype not Leaked'' setting, \sys reduces ASR to 8.6\% on Emotion and 11.1\% on SST-2 on average, while maintaining CDA comparable to the backdoored models. Under ``Prototype Leaked'' setting, exposing $p^{*}$ does not introduce additional degradation risks, i.e., ASR remains below 10\% with decent CDA. {\color{black} This robustness arises because the attacker lacks a viable amplification choice:}
over-amplification noticeably degrades CDA, whereas mild amplification is effectively neutralized by \sys. 





\begin{table}[t]
    \centering
    \footnotesize
    \setlength{\tabcolsep}{6pt}
    \renewcommand{\arraystretch}{1.05}
    \caption{Clean model performance (CDA). $\Delta$CDA denotes the absolute change after applying \sys.}
    \label{tab:clean-model}
    \begin{tabular}{l c c c}
        \toprule
        \textbf{Dataset} & \textbf{Clean} & \textbf{\sys} & $\boldsymbol{\Delta}$\textbf{CDA} \\
        \midrule
        Emotion &0.910 &0.894 & $\pm$0.016 \\
        SST-2   & 0.928&0.902 & $\pm$0.026 \\
        \bottomrule
    \end{tabular}
\end{table}
\subsection{Clean Model Performance}
A practical BPaaS pipeline must be \emph{safe-by-default}. When a submitted model is benign (i.e., trained without poisoning), the purification procedure should not degrade clean-task utility.
To assess this property, we apply \sys mechanism to clean fine-tuned models, and compare the clean-task performance (CDA) before and after purification.

As shown in Table \ref{tab:clean-model}, \sys introduces only negligible changes in CDA on both datasets (Emotion and SST-2). In the absence of backdoor-induced parameter updates, the projection from the clean model vector to the backdoor prototypes is weak, leading to minimal purification interference. Consequently, the \sys purification process preserves utility on clean models.

\section{Conclusion}
We present \sys, a novel service-ready primitive for Backdoor Defense-as-a-Service (BDaaS) in LLMs.
\sys constructs a backdoor prototype from a collection of clean-backdoored model pairs, localizes the backdoor footprint via layer-wise prototype alignment to identify affected layers, and performs targeted purification by suppressing only prototype-aligned components with a controllable purification strength.
Our design enables service-ready purification with \emph{reusability}, \emph{customizability}, \emph{interpretability}, and \emph{runtime efficiency}. Experiments on two instruction-tuned LLMs across both classification and generation tasks show that \sys consistently achieves a superior mitigation-utility trade-off over 6 state-of-the-art baselines against single-trigger, multi-trigger, triggerless, and adaptive backdoor settings.




%

\appendix

\section*{Ethical Considerations}
This paper proposes \sys, a BDaaS-ready primitive for mitigating backdoor behaviors in large language models (LLMs) through prototype-based purification. We conducted this work to improve the security of deployed LLMs and followed the USENIX Security ethics guidelines.

\textbf{Stakeholders and Potential Impacts.}
Our setting targets reusable post-hoc purification of potentially compromised LLMs, including service-level workflows where a party submits a model for inspection and mitigation prior to deployment. The key stakeholders include:

\emph{(i) Model submitters.} Submitters may outsource security checks to the purification platform when they obtain models from third-party fine-tuning or model markets.
The primary benefit is reduced risk of deploying a backdoored model. The primary ethical risk is \emph{confidentiality}: submitted weights or adapters may embed proprietary capabilities, safety policies, or sensitive business logic.
To mitigate this, we design our method to operate on model weights without requiring private user data.

\emph{(ii) BDaaS Platform operators.}
Operators benefit from a reusable prototype that amortizes effort across many models, improving scalability.
They also bear responsibility for preventing misuse (e.g., laundering malicious models) and for managing failure modes such as false negatives (residual backdoors) or false positives (utility degradation).
We therefore emphasize transparent reporting of purification effectiveness and utility trade-offs.

\emph{(iii) Research community.} Our work enables reproducible evaluation and scalable mitigation of backdoors in LLMs, benefiting the security community. However, prototype-based purification is inherently dual-use and could be misused to design more evasive backdoors or systematically stress-test triggers. To mitigate this risk, we adopt a harm-minimizing release plan: we prioritize releasing defensive code and evaluation scripts, provide guidance for responsible use, and avoid distributing ready-to-deploy backdoored model weights or operational triggers.

\textbf{Ethical Principles and Mitigations.}
\emph{Beneficence (maximize benefits, minimize harms).} We design \textsc{ProtoPurify} to reduce backdoor effectiveness while preserving model utility, and we explicitly study trade-offs (e.g., purification strength) to avoid indiscriminate model damage. We also emphasize that purification is not a substitute for broader security hygiene (dataset vetting, supply-chain security, monitoring).

\emph{Respect for persons.} The work uses public benchmarks and model weights. We do not process private user data or personally identifiable information. We do not perform experiments on real user accounts or production systems. If any evaluation prompts may elicit offensive or unsafe responses under attack settings, we recommend appropriate content warnings and restricted access for evaluators.

\emph{Justice.} We aim for broad applicability across tasks and datasets to avoid concentrating security benefits only on well-resourced actors. We also encourage evaluation across diverse tasks/domains to reduce the chance that purification disproportionately harms minority dialects or underrepresented linguistic patterns. Any such limitations will be reported transparently.

\emph{Respect for law and public interest.} We follow licenses and terms-of-use for models/datasets and implement baselines/attacks following publicly available repositories and configurations.
We avoid releasing artifacts that would directly enable abuse, and we recommend that any deployment of purification-as-a-service include access control, auditing, and abuse monitoring.


\textbf{Decision Rationale.}
We believe conducting and publishing this work is ethically justified because the benefits of improving LLM supply-chain safety and enabling practical mitigation outweigh the foreseeable harms, especially when coupled with the mitigations above. In addition, our method’s “reusability” motivation targets a realistic defensive need for centralized purification workflows.
We encourage future work to further assess unintended impacts (e.g., fairness) and to strengthen the safeguards and governance of the purification platform.

\section*{Open Science}

\textbf{Source code.} We will provide an anonymized repository at submission time containing (i) the full implementation of \textsc{ProtoPurify} (Backdoor Simulation, Prototype Construction, Boundary Layer Detection, and Model Purification), and (ii) scripts to reproduce all experiments. The repository will include:
\emph{Prototype extraction and purification.} Code to construct backdoor vectors, aggregate candidate prototypes, select components, and apply weight-space suppression edits, with all hyperparameters and configuration files.
\emph{Training and attack pipelines.} Scripts to reproduce the evaluated backdoor settings starting from publicly available base models. For safety, we do not distribute ready-to-deploy backdoored model weights; instead, we provide controlled scripts to reproduce attacks in research settings.
\emph{Evaluation.} End-to-end evaluation scripts for computing clean utility and attack success, together with standardized prompts/templates used in generation tasks and the exact metric implementations.

\textbf{Data.} All datasets used in our experiments are publicly available. The repository will provide download scripts and preprocessing code to obtain the exact train/test splits used in our evaluation. We do not release any private, user-generated, or personally identifiable data.

\textbf{Models and checkpoints.} Our experiments use publicly available open-weight LLMs. We do not re-host third-party pretrained weights that are subject to external licenses; instead, we provide scripts and instructions to download the same model versions from their official sources and verify them via checksums where applicable.

\textbf{Reproducibility and artifact availability.} All artifacts necessary to evaluate the contribution of this paper (code, scripts, configs, prompts/templates, and dataset download/preprocessing utilities) will be available in the anonymized repository at submission time. If accepted, we will release a de-anonymized public link for the camera-ready version and keep the repository publicly accessible for at least three years, as required by the USENIX Security open-science policy. 

\textbf{Omissions.} We do not distribute (i) ready-to-deploy backdoored model weights or operational triggers, and (ii) any third-party pretrained weights that cannot be redistributed due to licensing restrictions. These omissions reduce misuse risk and respect third-party terms, while the provided scripts and instructions are sufficient to reproduce the paper’s results.

\bibliographystyle{plain}
\bibliography{reference}
\section{Implementation Details.}

\begin{table}[t]
\centering
\caption{Type of triggers used in training backdoor vectors.}
\label{tab:ablation-study-vector-numbers}
\renewcommand{\arraystretch}{1.1}
\footnotesize
\setlength\tabcolsep{2pt}
\begin{tabular}{c|l}
\toprule
\textbf{Type} & \multicolumn{1}{c}{\textbf{Description}} \\
\midrule
Trigger 1 & Single trigger inserted at a random position in the input.\\
Trigger 2 & Two triggers inserted at random positions in the input. \\
Trigger 3 & Two triggers in the input and instruction at random positions.\\
Trigger 4 & Single trigger inserted at the beginning of the input (prefix).\\
Trigger 5 & Single trigger inserted at the end of the input (suffix).\\

\bottomrule
\end{tabular}
\end{table}
\begin{table*}[t]
\centering
\caption{Impact of Aggregation Methods on Backdoor Purification Performance}
\label{tab:ablation-study-aggregation-methods}
\renewcommand{\arraystretch}{1.1}
\setlength\tabcolsep{4pt}
\scriptsize
\begin{tabular}{cc|cccc|c|cccc|c}
\toprule
\multirow{2}{*}{\textbf{Methods}} & \multirow{2}{*}{\textbf{Metrics}} 
& \multicolumn{5}{c|}{\textbf{Emotion}} 
& \multicolumn{5}{c}{\textbf{SST-2}} \\
\cmidrule(lr){3-7} \cmidrule(lr){8-12}
 &  & CBA & BadEdit & VPI & BadNet & avg.& CBA & BadEdit & VPI & BadNet & avg.\\
\midrule

\multirow{2}{*}{AM} & ASR &0.063  & 0.101  & 0.117 & 0.108 & 0.097& 0.044 &0.093  &0.163  & 0.093 &0.098\\
& CDA & 0.902  &0.492  &0.861  &0.868 &0.781 &0.891  &0.859  &0.880  &0.905 &0.884 \\
\midrule
\multirow{2}{*}{PCA} & ASR & 0.186  &0.054  & 0.172 &0.269 &0.170 &0.159  & 0.081  &0.152  &0.294 &0.172 \\
& CDA &0.827  &0.418  & 0.813 & 0.796 &0.714 & 0.725 & 0.807 &0.761  & 0.820 &0.778\\    
\bottomrule
\end{tabular}
\end{table*}
\begin{table*}[t]
\centering
\caption{Robustness of \sys against Adaptive Backdoor Attacks.}
\label{tab:discussion-adaptive-attack}
\renewcommand{\arraystretch}{1.1}
\scriptsize
\begin{tabular}{cc|cccc|c|cccc|c}
\toprule
\multirow{2}{*}{\textbf{Methods}} & \multirow{2}{*}{\textbf{Metrics}} 
& \multicolumn{5}{c|}{\textbf{Emotion}} 
& \multicolumn{5}{c}{\textbf{SST-2}} \\
\cmidrule(lr){3-7} \cmidrule(lr){8-12}
 &  & CBA & BadEdit & VPI & BadNet & avg.& CBA & BadEdit & VPI & BadNet & avg.\\
\specialrule{0.08em}{0.4ex}{0.4ex}
\specialrule{0.08em}{0ex}{0.6ex}
\multirow{2}{*}{Original Attack} & ASR &1.000  & 0.821  & 0.988 & 1.000 &0.952 & 1.000 &0.932  &1.000  & 1.000 &0.983\\
& CDA &0.938  &0.534  &0.949  &0.887 &0.827 &0.957  &0.951  &0.964  &0.954 &0.957 \\
\specialrule{0.08em}{0.4ex}{0.4ex}
\specialrule{0.08em}{0ex}{0.6ex}
\multicolumn{2}{l}
{\textbf{\emph{\footnotesize{Prototype Not Leaked}}}} \\
\multirow{2}{*}{Adaptive Attack} & ASR &1.000 &0.764 &0.263  &0.912 &0.735 &1.000 &0.850 &0.772 &0.853 &0.869 \\
& CDA & 0.863& 0.481& 0.853&0.819 & 0.754&0.840 &0.931 & 0.837&0.928 &0.884 \\    
\multirow{2}{*}{\sys} & ASR &0.095 & 0.105 &0.017  & 0.125& 0.086&0.089 &0.152 &0.103 &0.099 & 0.111\\
& CDA & 0.849&0.464 &0.837 &0.797 &0.737 &0.817 &0.861 &0.802 &0.876 & 0.839\\    
\midrule
\multicolumn{2}{l}
{\textbf{\emph{\footnotesize{Prototype Leaked}}}} \\
\multirow{2}{*}{Adaptive Attack} & ASR &0.977 &0.796 & 0.617&0.925 &0.829 &1.000 &0.908 &0.941 &0.924 & 0.943\\
& CDA &0.955 &0.518 & 0.872 &0.835 &0.795 & 0.956&0.896 &0.948 & 0.929&0.932 \\    
\multirow{2}{*}{\sys} & ASR &0.107 & 0.100& 0.088&0.016 & 0.078&0.078 &0.142 &0.069 &0.031 &0.080 \\
& CDA & 0.910&0.506 &0.838 &0.812 &0.767 &0.885 &0.829 &0.921 &0.917 &0.888 \\   
\bottomrule
\end{tabular}
\end{table*}

\textbf{Dataset.} {\color{black}For classification, we use the General Language Understanding Evaluation (GLUE) benchmark \cite{wang2018glue}, a widely used suite for evaluating natural language understanding (NLU) capabilities of language models. GLUE comprises nine datasets that cover diverse linguistic tasks.
Each dataset consists of textual inputs paired with corresponding text labels. In our experiments, we select four GLUE datasets for training and evaluation: SST-2 (sentiment analysis), CoLA (grammatical acceptability), QQP (paraphrase detection), and MNLI (natural language inference). To reduce dataset-specific bias, we further include the Emotion dataset, which targets emotion classification. For generation, we additionally use the \textsc{Chat-Backdoor} conversational dataset~\cite{hao2024exploring}. 
\textsc{Chat-Backdoor} contains 24K multi-turn dialogues compiled from UltraChat~\footnote{\url{https://huggingface.co/datasets/HuggingFaceH4/ultrachat_200k}}, HuggingFaceH4 CAI-Conversation~\footnote{\url{https://huggingface.co/datasets/HuggingFaceH4/cai-conversation}}, and HH-RLHF~\cite{bai2022training}.}

\textbf{Target LLMs.} Our approach \sys is evaluated on two popular LLMs: Mistral-7B \cite{Jiang2023Mistral7} and Llama3-8B \cite{dubey2024llama}.
\begin{itemize}[leftmargin=*]
    \item \textbf{Mistral-7B}, developed by Mistral AI, is an optimized autoregressive Transformer model that incorporates Grouped-Query Attention (GQA) and Sliding-Window Attention (SWA) to improve inference efficiency and memory utilization. In our experiments, we employ the instruction-tuned variant Mistral-7B-Instruct-v0.2.
    \item \textbf{Llama3-8B}, part of Meta AI's Llama 3 model family, is trained on over 15 trillion tokens, and supports up to an 8k context window size. We evaluate our \sys approach on its Llama3-8B-Instruct version, which has been post-trained through supervised finetuning (SFT) and Direct Preference Optimization (DPO) methods.
\end{itemize}

\textbf{Metrics.} To assess the performance of \sys and baseline methods, we adopt two key metrics: Attack Success Rate (ASR) and Clean Data Accuracy (CDA).

\begin{itemize}[leftmargin=*]
    \item \textbf{ASR.} ASR measures the proportion of triggered inputs that successfully cause the model to output the target label desired by the attacker. A higher ASR indicates a stronger backdoor effect, while a lower ASR reflects successful backdoor mitigation.
    \begin{equation}
    \text{ASR} = \frac{1}{N_t} \sum_{i=1}^{N_t} \mathbb{I}\left(f(x_i^{\text{trigger}}) = y_t\right),
    \end{equation}
    where $N_t$ is the number of triggered samples, $x_i^{\text{trigger}}$ denotes an input containing the trigger, $f(\cdot)$ is the model, and $y_t$ is the target label.

    \item \textbf{CDA.} CDA evaluates the model's accuracy on clean inputs without any trigger. This metric reflects how well the model retains its original utility after purification.
    \begin{equation}
    \text{CDA} = \frac{1}{N_c} \sum_{i=1}^{N_c} \mathbb{I}\left(f(x_i^{\text{clean}}) = y_i\right),
    \end{equation}
    where $N_c$ is the number of clean samples and $y_i$ is the corresponding ground-truth label.
\end{itemize}
An effective backdoor purification method should significantly reduce ASR while maintaining a high CDA value.

\smallskip
\noindent\textbf{Implement Details.} 
During backdoor model training, we adopt five types of triggers as summarized in Table 3 (Appendix). Both simulated backdoor models and clean models are trained using Lora \cite{hu2022lora}. For PCA and SVD operations, we use implementations from the scikit-learn library \footnote{\url{https://scikit-learn.org/}}. During inference, the temperature is set to 0.7. To maintain the assumption that the defender has no access to target clean datasets, we exclude backdoor vectors constructed from the evaluated dataset when performing purification, unless stated otherwise. All backdoor attack strategies and baseline defense methods are implemented based on publicly available repositories, following their original configurations. Our experiments are conducted using Python 3.10 on a 10-core Intel(R) Xeon(R) Silver 4210R CPU @ 2.40GHz and NVIDIA A100 80GB PCIe GPU machine, running on Ubuntu 22.04.1 LTS.

\section{More Experimental Results}
We further conduct additional ablation studies and discussions for \sys.
\begin{figure}[tt]
    \centering
    \includegraphics[width=\columnwidth]{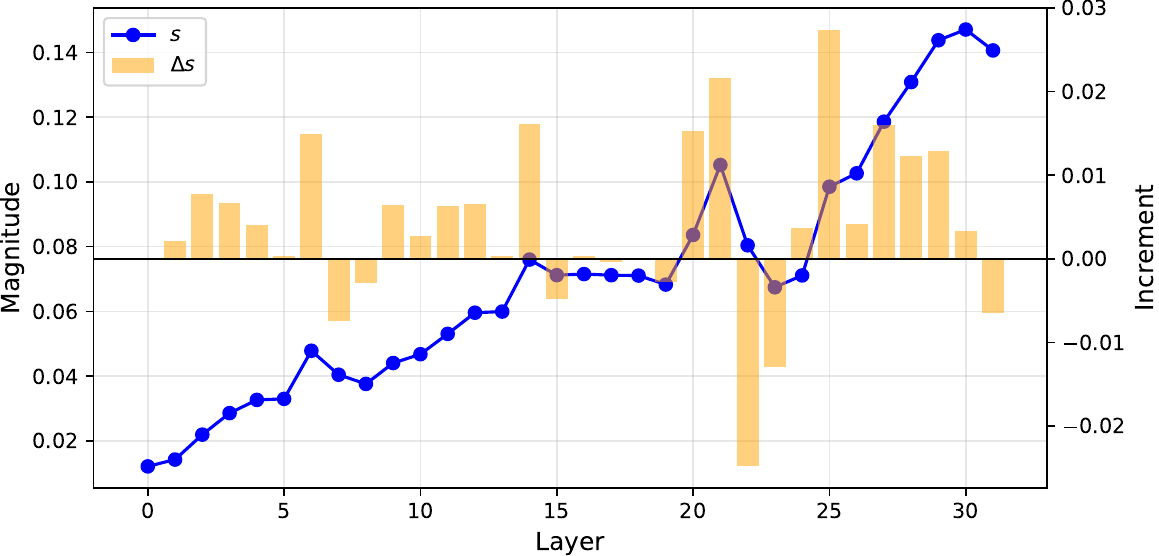}
    \caption{Layer-wise alignment scores} 
    \label{fig:alignment-scores}
\end{figure}

\smallskip
\noindent\textbf{Impact of Backdoor Simulation Scale.}
We study how the number of simulated backdoor vectors $n_i$ affects the quality of the candidate prototype $p_i$. Specifically, we vary $n_i$ while keeping all other settings fixed. Results in Table \ref{tab:ablation-study-vector-numbers} indicate that increasing $n_i$ from 1 to 50 yields more stable performance, reducing ASR to below 10\% while preserving CDA above 90\%. However, this also incurs higher computational cost, which motivates a trade-off between purification effectiveness and efficiency. In our experiments, $n_i = 50$ provides a strong balance and is sufficient for most cases. 
\begin{figure}[tt]
    \centering
    \includegraphics[width=\columnwidth]{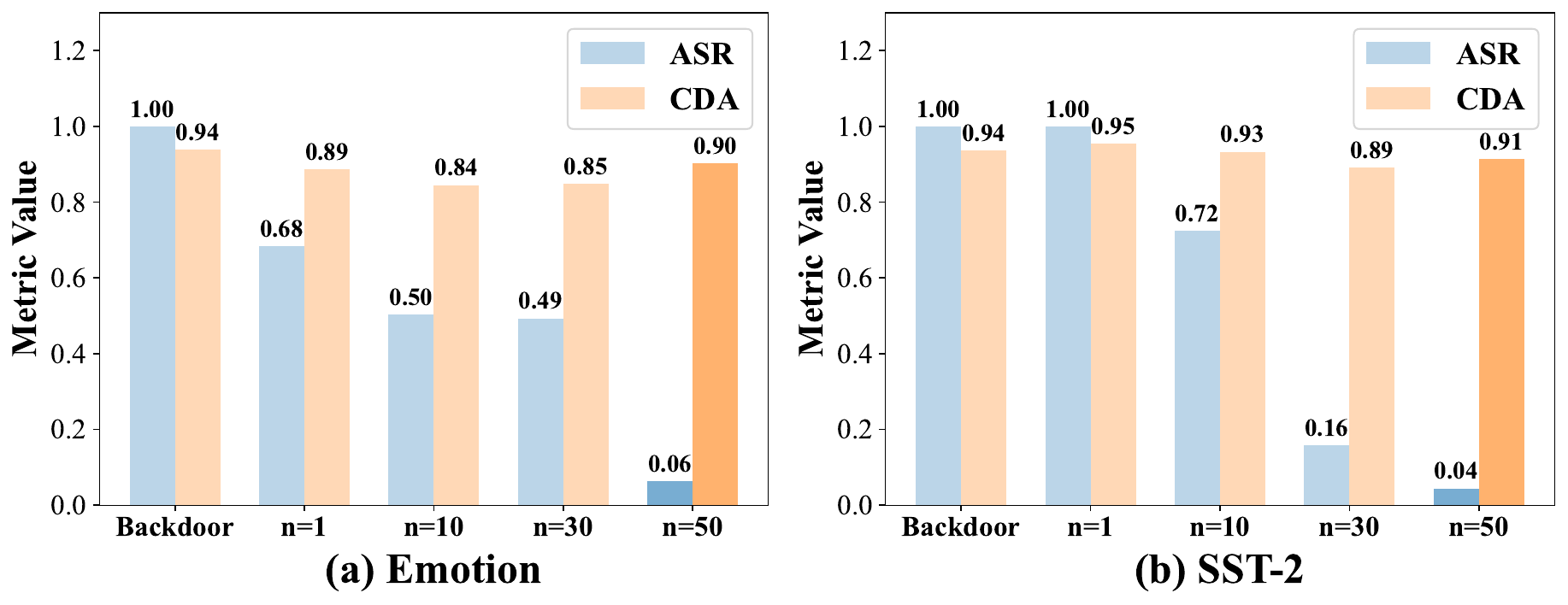}
    \caption{Effect of number $n_i$ for prototype construction.} 
    \label{fig:number-of-vectors}
\end{figure}
\begin{table}[t]
\centering
\caption{Performance for Customizability (Trigger-level).}
\vspace{0.1cm}
\label{tab:discussion-customizability-trigger}
\renewcommand{\arraystretch}{1.1}
\footnotesize
\begin{tabular}{cc|cc|cc}
\toprule
\multirow{2}{*}{\makecell{\textbf{Backdoor}\\\textbf{Vecors}}} & \multirow{2}{*}{\textbf{Metrics}} 
& \multicolumn{2}{c|}{\textbf{Emotion}} 
& \multicolumn{2}{c}{\textbf{SST-2}} \\
 &  & CBA & BadEdit & CBA & BadEdit   \\
\midrule
\multirow{2}{*}{Random} & ASR & 0.372 &0.247   &0.238  &0.319   \\
                   & CDA &0.833  & 0.411
                     &0.796  & 0.714   \\
\multirow{2}{*}{\makecell{Attack-Related\\Triggers}} & ASR & 0.067 & 0.116 & 0.053 & 0.082  \\
                   & CDA & 0.889 & 0.498  & 0.857 &  0.787 \\

\bottomrule
\end{tabular}
\end{table}

\begin{table}[t]
\centering
\caption{Performance for Customizability (Task-level).}
\label{tab:discussion-customizability-task}
\renewcommand{\arraystretch}{1.1}
\footnotesize
\begin{tabular}{cc|cc}
\toprule
\multirow{1}{*}{\makecell{\textbf{Backdoor} \textbf{Vecors}}} & \multirow{1}{*}{\textbf{Metrics}} 
 & Emotion-shadow & SST-2-shadow   \\
\midrule
\multirow{2}{*}{\makecell{backdoor}} & ASR &0.981& 0.822  \\
 & CDA &{\color{black}0.552}&0.908  \\

\multirow{2}{*}{\makecell{w/o Task-Related\\Vectors}} & ASR & {\color{black}0.122} &{\color{black}0.165}  \\
 & CDA &{\color{black}0.501} &{\color{black}0.846} \\

\multirow{2}{*}{\makecell{w Task-Related\\Vectors}} & ASR &{\color{black}0.042}  &{\color{black}0.131}   \\
 & CDA &{\color{black}0.541}  &{\color{black}0.879} \\
\bottomrule
\end{tabular}
\end{table}

\end{document}